\documentclass[prl,aps,floatfix,twocolumn,superscriptaddress]{revtex4-2}

\usepackage[english]{babel}
\usepackage{amsmath}
\usepackage{graphicx}
\usepackage{float}
\usepackage{bm}
\usepackage{multirow}
\usepackage{dcolumn}
\usepackage{color}
\usepackage{hyperref}

\newcommand{\icm}{\ensuremath{~\textrm{cm}^{-1}}}% % cm-1

\begin{document}

\title{\textbf{Spectroscopic signatures of magnetization-induced band renormalization and strong spin-charge-lattice coupling in EuZn$_2$As$_2$}}

\author{Zhiyu Liao}
\thanks{These authors contributed equally to this work.}
\author{Boxuan Li}
\thanks{These authors contributed equally to this work.}
\author{Shaohui Yi}
\affiliation{Beijing National Laboratory for Condensed Matter Physics, Institute of Physics, Chinese Academy of Sciences, P.O. Box 603, Beijing 100190, China}
\affiliation{School of Physical Sciences, University of Chinese Academy of Sciences, Beijing 100049, China}

\author{Lincong Zheng}
\author{Yubiao Wu}
\affiliation{Beijing National Laboratory for Condensed Matter Physics, Institute of Physics, Chinese Academy of Sciences, P.O. Box 603, Beijing 100190, China}

\author{Enkui Yi}
\affiliation{State Key Laboratory of Optoelectronic Materials and Technologies, School of Physics, Sun Yat-Sen University, Guangzhou, Guangdong 510275, China}

\author{Premysl Marsik}
\affiliation{University of Fribourg, Department of Physics and Fribourg Center for Nanomaterials, Chemin du Mus\'{e}e 3, CH-1700 Fribourg, Switzerland}

\author{Bing Shen}
\email[]{shenbing@mail.sysu.edu.cn}
\affiliation{State Key Laboratory of Optoelectronic Materials and Technologies, School of Physics, Sun Yat-Sen University, Guangzhou, Guangdong 510275, China}

\author{Hongming Weng}
\email[]{hmweng@iphy.ac.cn}
\affiliation{Beijing National Laboratory for Condensed Matter Physics, Institute of Physics, Chinese Academy of Sciences, P.O. Box 603, Beijing 100190, China}
\affiliation{School of Physical Sciences, University of Chinese Academy of Sciences, Beijing 100049, China}

\author{Bing Xu}
\email[]{bingxu@iphy.ac.cn}
\affiliation{Beijing National Laboratory for Condensed Matter Physics, Institute of Physics, Chinese Academy of Sciences, P.O. Box 603, Beijing 100190, China}
\affiliation{School of Physical Sciences, University of Chinese Academy of Sciences, Beijing 100049, China}

\author{Xianggang Qiu}
\affiliation{Beijing National Laboratory for Condensed Matter Physics, Institute of Physics, Chinese Academy of Sciences, P.O. Box 603, Beijing 100190, China}
\affiliation{School of Physical Sciences, University of Chinese Academy of Sciences, Beijing 100049, China}

\author{Christian Bernhard}
\affiliation{University of Fribourg, Department of Physics and Fribourg Center for Nanomaterials, Chemin du Mus\'{e}e 3, CH-1700 Fribourg, Switzerland}

%%%%%%%%%%%%%%%%%%%%%%%%%%%%%%%%%%%%
%
% Abstract

\begin{abstract}
We report an infrared spectroscopy study of the antiferromagnetic (AFM) insulator EuZn$_2$As$_2$ over a broad frequency range, spanning temperatures both above and below the AFM transition $T_{\rm N} \simeq$ 20 K. The optical response reveals an insulating behavior, featuring two prominent infrared-active phonon modes at around 95 and 190 cm$^{-1}$, and two subtle absorption peaks at around 130 ($\alpha$ peak) and 2\,700 cm$^{-1}$ ($\beta$ peak), along with a strong absorption edge rising around 9\,000 cm$^{-1}$ ($\gamma$ peak). Significantly, the temperature-dependent changes in these peaks show noticeable anomalies across the AFM transition, particularly the emergence of the $\alpha$ peak and an unusual redshift of the $\gamma$ peak, suggesting a strong interaction between the charge excitations and the AFM order. Band structure calculations reveal that these anomalies arise from magnetization-induced band renormalizations, including shifts and foldings. Additionally, both phonon modes feature asymmetric Fano line shapes at low temperatures, with the 95 cm$^{-1}$ phonon mode exhibiting strong coupling to the fluctuations of Eu spins. These findings highlight a complex interplay of spin, charge, and lattice degrees of freedom in EuZn$_2$As$_2$.
\end{abstract}

\maketitle

%%%%%%%%%%%%%%%%%%%%%%%%%%%%%%%%%%%%%%%%%%%%%%%%%%%%%%%%%%%%%%%%%%%%%%%%%%%%%%%
%
% Introduction

Recently, europium-based 122 pnictides EuM$_2$X$_2$ (M = Zn, In, Cd; X = P, As, Sb) have attracted significant attention as promising candidates for intrinsic magnetic topological materials~\cite{yan2017topological,bernevig2022progress,wang2023intrinsic,chen2024recent}. In these compounds, the interplay between the Eu-$4f$ magnetic moment and the electronic band structure gives rise to a variety of fascinating quantum states, making these materials particularly intriguing for both fundamental research and potential spintronic applications. For instance, in EuCd$_2$As$_2$~\cite{ma2019spin,soh2019ideal,wang2019single,ma2020emergence,taddei2022single}, the highly-tunable Eu moments can induce nontrivial topological phases, such as Dirac semimetal, Weyl semimetal, and higher-order topological insulator. In EuIn$_2$As$_2$ and EuSn$_2$As$_2$~\cite{xu2019higher,li2019dirac}, the magnetic ground states play a key role in forming the axion insulator states, highlighting the importance of magnetic order in shaping the topological properties. Additionally, Eu-based pnictides are well-known for exhibiting colossal magnetoresistance  effect~\cite{jiang2006colossal,Zhang2020PRB,wang2021colossal,yan2022field,li2021magnetic,du2022consecutive,zhang2023electronic,krebber2023colossal}.
Therefore, the strong spin-charge interaction in Eu-based materials can provide an ideal platform for manipulating band topology and exploring novel physical properties.

\begin{figure*}[tb]
\includegraphics[width=2\columnwidth]{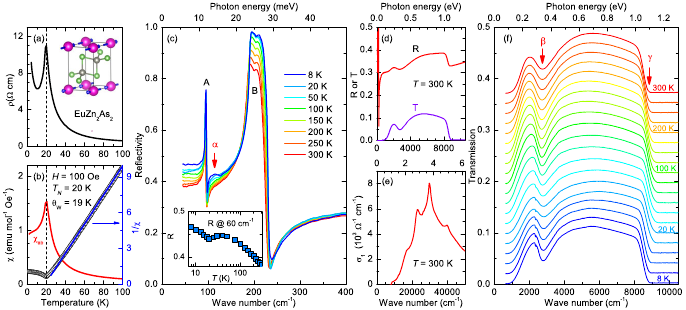}
\caption{(a) Temperature-dependent resistivity $\rho(T)$ with the current applied along the \emph{ab}-plane for EuZn$_2$As$_2$. Inset shows the schematic crystal structure and magnetic configuration of Eu atoms. (b) Temperature dependence of magnetic susceptibility (red line, left axis) and the inverse of magnetic susceptibility (black circles, right axis). The magnetic field $H$ is applied parallel to the \emph{ab}-plane. Solid blue line is the fitting curve by the Curie-Weiss law. (c) Temperature-dependent in-plane reflectivity spectra of EuZn$_2$As$_2$ in the far-infrared region. Inset: temperature-dependent reflectivity at 60 cm$^{-1}$. (d) The reflectivity and transmission spectra at 300 K up to 10\,000 cm$^{-1}$. (e) Optical conductivity spectrum measured by ellipsometry up to 50\,000 cm$^{-1}$ at 300 K. (f) Temperature dependence of the in-plane transmission, with the data offset for clarity.}
\label{fig1}
\end{figure*}

In contrast to the metallic hole-doped nature of the aforementioned several Eu-122 compounds, the newly discovered EuZn$_2$As$_2$ exhibits an antiferromagnetic (AFM) insulator ground state with $T_{\rm N} \simeq 20$ K~\cite{wang2022anisotropy,blawat2022unusual,bukowski2022canted,yi2023topological}. This compound has the same trigonal CaAl$_2$Si$_2$-type crystal structure (space group $P\bar{3}m1$, No. 164) and A-type AFM order as EuCd$_2$As$_2$. In this structure, the Zn$_2$As$_2$ bilayers are separated by the triangular Eu layers, where the Eu local spins align ferromagnetically within the ab-plane and point along the a/b-axis, while they order antiferromagnetically between adjacent Eu layers, as illustrated in the inset of Figure~\ref{fig1}(a). Transport measurements in EuZn$_2$As$_2$ have revealed dominant ferromagnetic fluctuations above $T_{\rm N}$ that result in an anomalous Hall effect and large negative magnetoresistance~\cite{yi2023topological,luo2023colossal}. First-principles calculations have predicted a topologically trivial insulator in the AFM phase, and a Weyl semimetal phase under ferromagnetic order~\cite{wang2022anisotropy,luo2023colossal}. Quantum oscillation measurements have unveiled a complex hole band shape with non-trivial topological features~\cite{blawat2023quantum}. Despite these intriguing findings, investigations into the underlying interplay between magnetism and the electronic degrees of freedom are scarce, particularly regarding how AFM order influences the topological properties and electronic structure of this material.

In this Letter, we present an infrared spectroscopy study to explore the electron and lattice dynamics in EuZn$_2$As$_2$. We examine the temperature-dependent optical response and, in particular, its changes at the onset of the magnetic order. Our optical results reveal that the AFM transition induces significant anomalies in both the charge and lattice excitations, providing valuable insights into the interplay between the charge, spin, and lattice degrees of freedom in this system. Additional band calculations show that the AFM order leads to a strong renormalization of the bands and a nearly closed bulk energy gap, placing EuZn$_2$As$_2$ near a topological phase transition.

\begin{figure*}[tb]
\includegraphics[width=2\columnwidth]{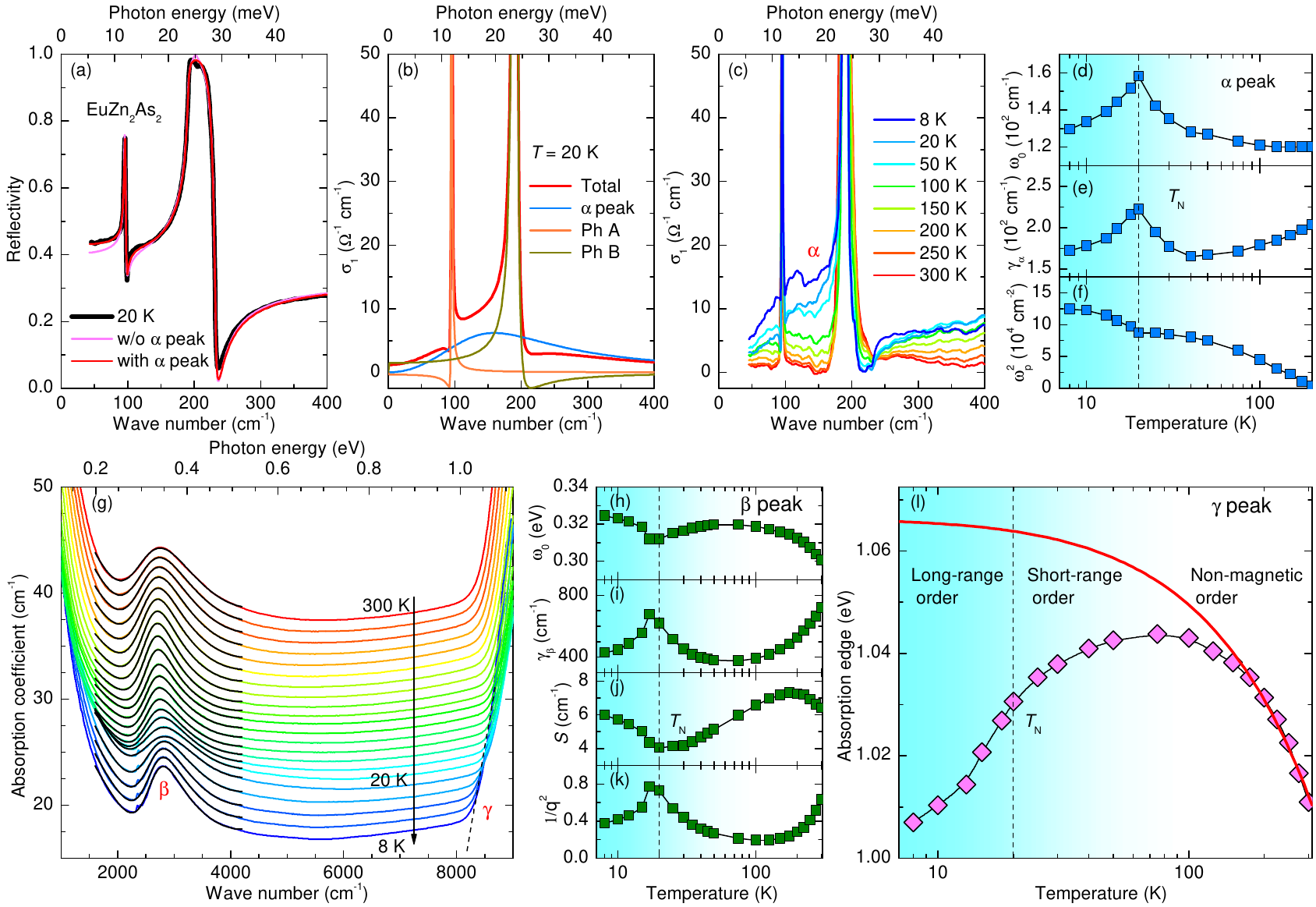}
\caption{(a) Drude-Lorentz fitting of the reflectivity at 20 K. (b) Decomposition of the corresponding real part of the optical conductivity $\sigma_1(\omega)$ at 20 K. (c) Temperature dependence of the $\sigma_1(\omega)$ spectra obtained from Kramers-Kronig analysis of reflectivity in the far-infrared range. (d--f) Temperature dependence of the resonance frequency $\omega_0$, linewidth $\gamma_\alpha$, and oscillator intensity $\omega^2_p$ for $\alpha$ peak. (g) Temperature dependence of the absorption coefficient. (h--k) Temperature dependence of the resonance frequency $\omega_0$, linewidth $\gamma_\beta$, and oscillator strength $S$ and Fano parameter $1/q^2$ for the $\beta$ absorption peak. (l) Temperature dependence of the energy gap edge of $\gamma$ peak.}
\label{fig2}
\end{figure*}

Sample synthesis, experimental methods, and details of the DFT calculations are provided in the Supplemental Materials~\footnotemark[1].

Figure~\ref{fig1}(a) displays the temperature ($T$) dependence of the resistivity $\rho(T)$ of EuZn$_2$As$_2$, which increases with decreasing $T$, suggesting a typical insulating behavior. In addition, a sharp resistivity peak around $T_{\rm N}$ = 20 K coincides with a corresponding peak in the $T$-dependent magnetic susceptibility $\chi(T)$, shown in Figure~\ref{fig1}(b). This feature is attributed to the AFM phase transition. Furthermore, the inverse susceptibility $1/\chi$ above $T_{\rm N}$ is well described by the Curie-Weiss law $ \chi(T) = C/(T-\theta_{\rm W}) $, where $\theta_{\rm W}$ is the Weiss temperature and $C$ is Curie constant. The positive value of $\theta_{\rm W}$ = 19 K indicates the presence of dominant ferromagnetic correlations between the Eu ions, despite the AFM ground state. The deduced effective magnetic moment $\mu_{\rm eff} =$ 8.08 $\mu_{\rm B}$ ($\mu_{\rm B}$ is the Bohr magneton) is in close agreement with the theoretical value of 7.94 $\mu_{\rm B}$ for the localized moment $4f^7$.

\begin{figure*}[tb]
\includegraphics[width=1.36\columnwidth]{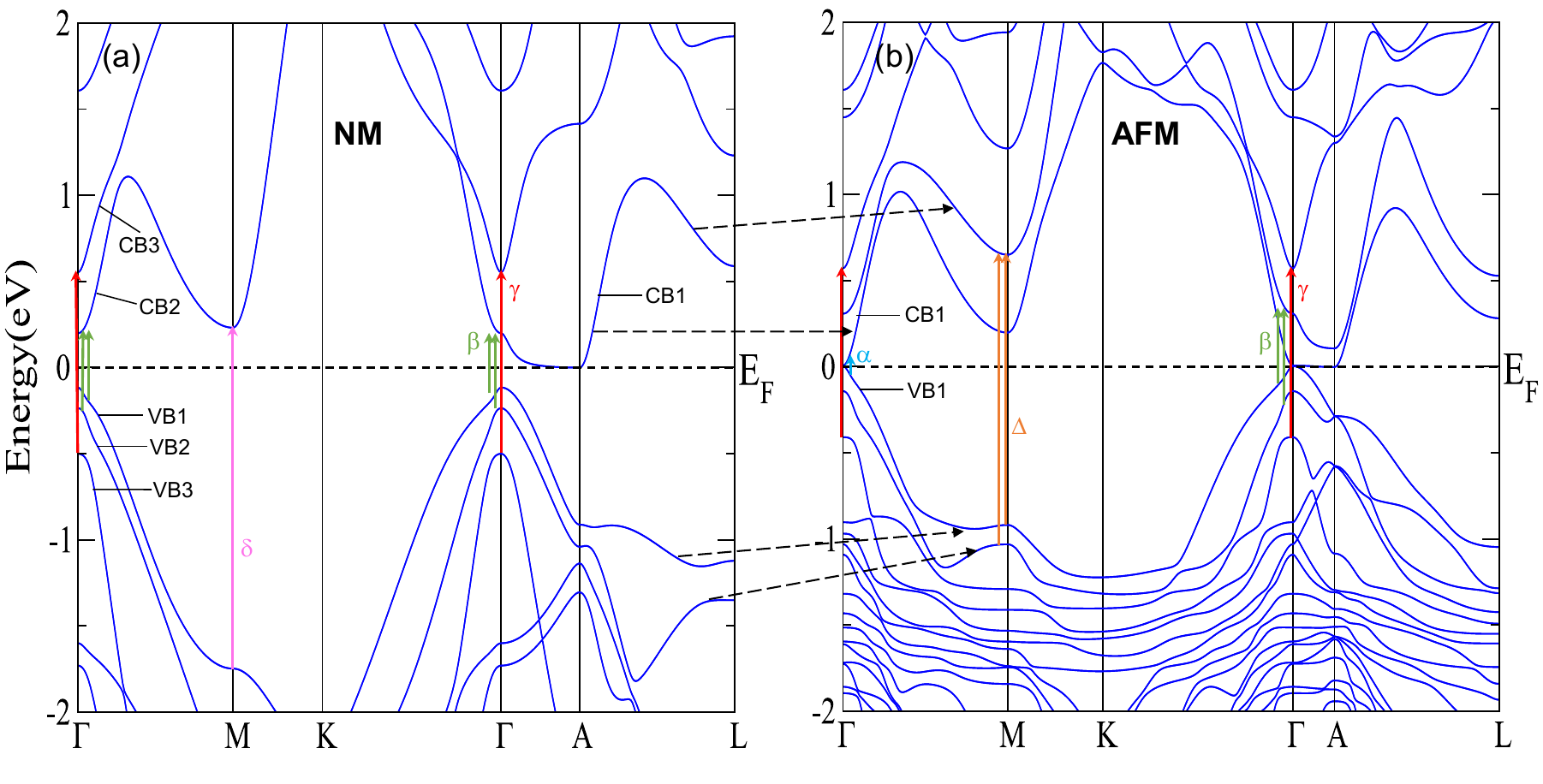}
\includegraphics[width=0.64\columnwidth]{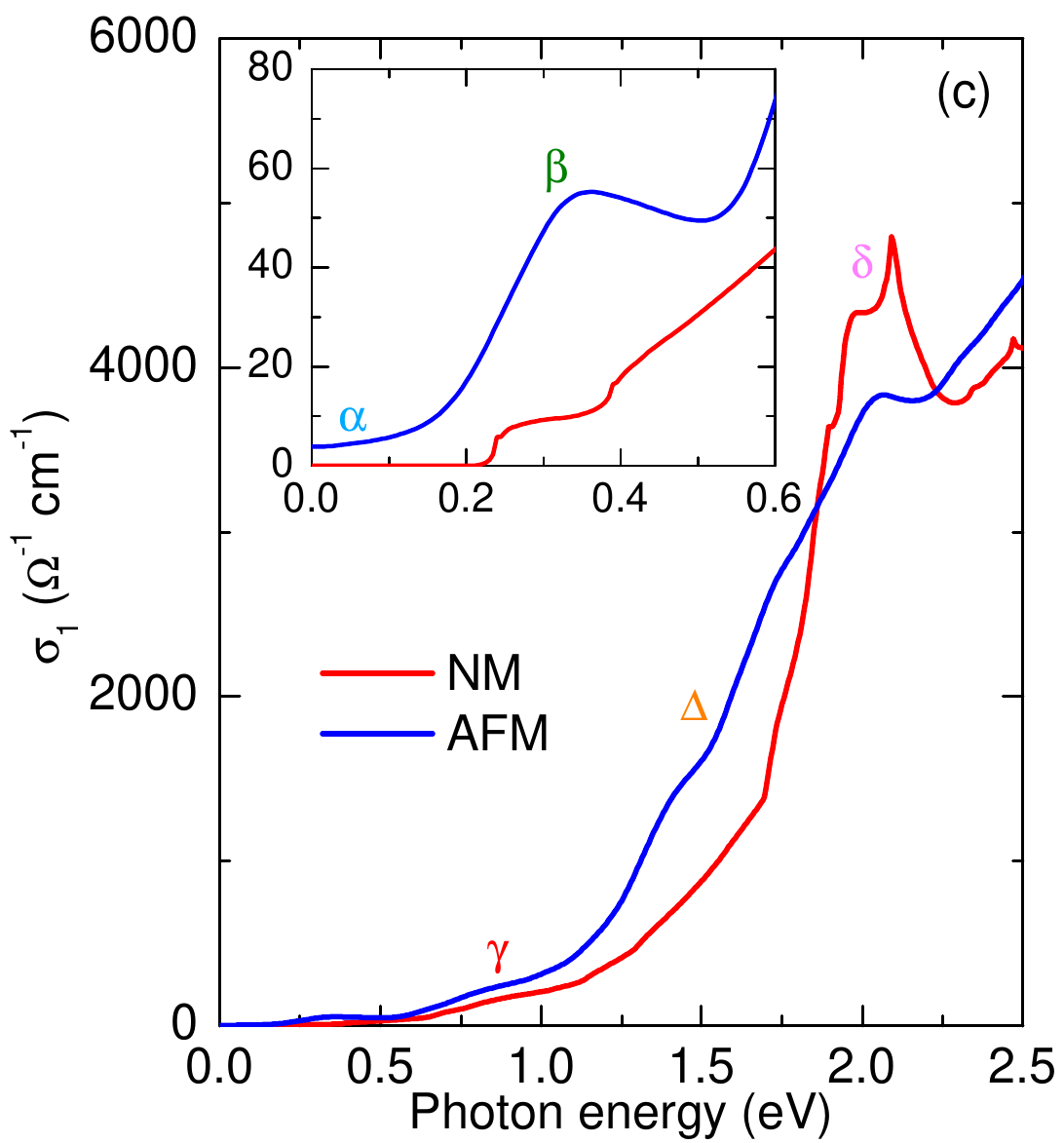}
\caption{(a) and (b) Calculations of the electronic band structure in nonmagnetic (NM) and antiferromagnetic (AFM) phases, respectively. (c) DFT calculations of $\sigma_1(\omega)$ in the NM and AFM phases up to 2.5 eV. Inset shows a zoom-in view of the spectra below 0.6 eV.}
\label{fig3}
\end{figure*}

Figure~\ref{fig1}(c) shows the in-plane reflectivity $R(\omega)$ of EuZn$_2$As$_2$ in the far-infrared region at several selected temperatures. The dominant features are two sharp peaks at 95 and 190 cm$^{-1}$ (labeled as A and B), which are attributed to infrared-active phonons with $E_u$ symmetry, reflecting the insulating nature of EuZn$_2$As$_2$. In addition, at low temperatures, a weak feature appears around 130 cm$^{-1}$ (labeled as $\alpha$), along with an increase in $R(\omega)$ in the low-frequency limit. This temperature dependence is highlighted in the inset, where the increasing trend of $R(\omega = 60\icm)$ with decreasing $T$ is interrupted by the formation of a dip near $T_{\rm N}$, indicating that the AFM order has a significant effect on the charge response.

In the high-frequency region, as shown in Fig.\ref{fig1}(d), the reflectivity remains nearly constant, but exhibits a steplike drop around 9\,000\icm, suggesting that the insulating sample is transparent. The transmittance measurements confirm this observation, since they show significant transmission between 1\,000 and 9\,000\icm. The step feature around 9\,000\icm\ corresponds to the absorption edge of the direct band gap. Furthermore, we obtained the optical conductivity at room temperature through ellipsometry measurements, as shown in Fig.\ref{fig1}(e). It reveals an absorption onset around 9\,000\icm\  (labeled as $\gamma$), followed by a series of sharp peaks at 17\,000, 23\,000, 30\,000, and 40\,000\icm, which are attributed to high-energy interband transitions. To investigate the detailed $T$ evolution of the insulating band gap, we measured transmission spectra at various temperatures, as shown in Fig.~\ref{fig1}(f). In the mid-infrared region, the sample is transparent with a transmission of about 10\%. A weak absorption feature exists around 2\,700\icm\ (labeled as $\beta$). In particular, as the temperature decreases, the absorption edge $\gamma$ first shows a blueshift and then an anomalous redshift, indicating an anomalous change in the electronic structure.

In order to obtain the optical conductivity $\sigma_1(\omega)$, we fit the measured $R(\omega)$ spectra at low energy using the Drude-Lorentz model. Details of the modeling are provided in the Supplemental Materials~\footnotemark[1]. Figure~\ref{fig2}(a) displays the fitting results of $R(\omega)$ at 20 K. Firstly, two Fano oscillators are used to fit the two phonon modes, where the measured $R(\omega)$ can be almost reproduced, except for the $\alpha$ peak feature and a lower reflectivity value in the zero-frequency limit. To improve the fit, an additional Lorentz oscillator is included to capture the $\alpha$ peak feature. With this adjustment, the fitting result agrees much better with the measured $R(\omega)$. Figure~\ref{fig2}(b) shows the corresponding optical conductivity, which consists of two phonon modes with asymmetric line shapes and the $\alpha$ peak centered around 130 cm$^{-1}$. No Drude response appears in the $\sigma_1(\omega)$ spectra, in line with our resistivity measurement. The $\sigma_1(\omega)$ spectra are also derived via the Kramers-Kronig analysis of $R(\omega)$ in the far-infrared range. As shown in Fig.~\ref{fig2}(c), $\sigma_1(\omega)$ near 130\icm\ gradually enhances upon cooling, indicating the emergence of the $\alpha$ component at low temperatures, in agreement with the modeled $\sigma_1(\omega)$. Figs.~\ref{fig2}(d--f) display the temperature dependence of the fitting parameters for the $\alpha$ peak. The parameters $\omega_{0}$, $\gamma_\alpha$, and $\omega_{p}$ all show distinct anomalies around $T_{\rm N}$, demonstrating a strong interplay between the spin and charge degrees of freedom.

Next, we examine the $T$-evolution of the charge excitations in the high-energy region. Figure~\ref{fig2}(g) shows the $T$ dependence of the absorption coefficient, $\alpha(\omega)=(1/d){\rm ln}[1/t(\omega)]$, where $d=1.3$ mm is the thickness of the sample and $t(\omega)$ is the transmission spectrum. The absorption peak ($\beta$) around 2\,700\icm\ exhibits an asymmetric line shape. As indicated by the thin black line, this peak is fitted using a Fano line shape with a quadratic background, $\alpha(\omega) = S\left[\frac{q^2 + 2qz -1}{q^2 (1 + z^2)}\right]$, where $z = (\omega-\omega_0)/\gamma_{\beta}$, and $\omega_0$, $\gamma_{\beta}$ and $S$ represent the frequency, linewidth, and strength of the $\beta$ peak, respectively. The asymmetric profile is determined by the Fano parameter $1/q^2$. The $T$-dependent fitting parameters are shown in Figs.~\ref{fig2}(h--k), all of which display anomalies around $T_{\rm N}$, signaling a close correlation between the charge excitations and the AFM order. Figure~\ref{fig2}(l) shows the temperature dependence of the absorption edge ($\gamma$), extracted by linearly extrapolating the data to the horizontal axis. In general, the variation of the energy gap with $T$ in semiconductors follows an empirical formula: $E_g(T)=E_0-aT^2/(T+b)$, where $E_0$ is the energy gap at 0 K, and $a$ and $b$ are constants~\cite{varshni1967temperature}. Since the absorption edge provides an estimate of the energy gap, we apply this formula to fit the $T$ dependence of the absorption edge. As shown by the red line in Figure~\ref{fig2}(l), the position of the absorption edge deviates clearly from the expected behavior below 100 K. At the lowest measured temperature, the energy gap decreases by approximately 60 meV compared to the expected value. Note that this empirical formula is satisfied only when the temperature-dependent lattice dilation and electron-lattice interactions are considered. Therefore, the suppression of the energy gap at low temperatures is due to additional effects, like the interplay between the charge and spin degrees of freedom.

To figure out the impact of AFM order on the charge response, we performed DFT calculations of the electronic band structure in both nonmagnetic (NM) and AFM phases. In the NM phase, shown in Fig.~\ref{fig3}(a), the energy gap between valence bands (VB1, VB2) and the conduction band (CB2) at the $\Gamma$ point is approximately 0.3 eV. This gap corresponds to the absorption peak $\beta$, while the absorption edge around 9\,000\icm\ is attributed to the interband transitions between VB3 and CB3. In the AFM phase, shown in Fig.~\ref{fig3}(b), there is a band reconstruction caused by the band folding along $k_z$ due to the unit cell doubling. The conduction band CB1 along the A-L direction folds to the $\Gamma$-M direction. Additionally, the Eu $4f$ orbital forms flat bands located at around 1.5--2.0 eV below the Fermi level. The band repulsion between the Eu-$4f$ bands and the broad As bands shifts the occupied valence bands upward, reducing the gap at the $\Gamma$ point. The band shift also leads to a decrease in the energy gap associated with the absorption edge $\gamma$. We also calculated the optical conductivity $\sigma_1(\omega)$ in the NM and AFM phases. As shown in Fig.~\ref{fig3}(c), the calculated $\sigma_1(\omega)$ exhibits a prominent absorption peak near 2 eV (labeled as $\delta$). Below this energy, the AFM $\sigma_1(\omega)$ shows higher spectral weight due to new excitations arising from AFM-induced band folding, such as the interband transitions denoted as $\Delta$ in Fig.~\ref{fig3}. In the low-energy region, as highlighted in the inset, both the NM and AFM calculations show considerable excitations at around 0.3 eV associated with the $\beta$ peak, while the absorption edge $\gamma$ corresponds to the sharp rise of the $\sigma_1(\omega)$ due to the strong interband absorptions. Below 0.2 eV, the NM result shows a fully open gap, while in the AFM result, the $\alpha$ absorption manifests itself as a very small Drude peak centered at zero frequency, as the energy gap is nearly closed and the Fermi level is located shallowly in the valence band. However, in our experiment, it appears as an absorption peak at a finite frequency. This difference arises because the strong spin-carrier coupling in this system would lead to carrier localization, as previously discussed and observed in EuCd$_2$P$_2$~\cite{Homes2023}.

\begin{figure}[b]
\includegraphics[width=\columnwidth]{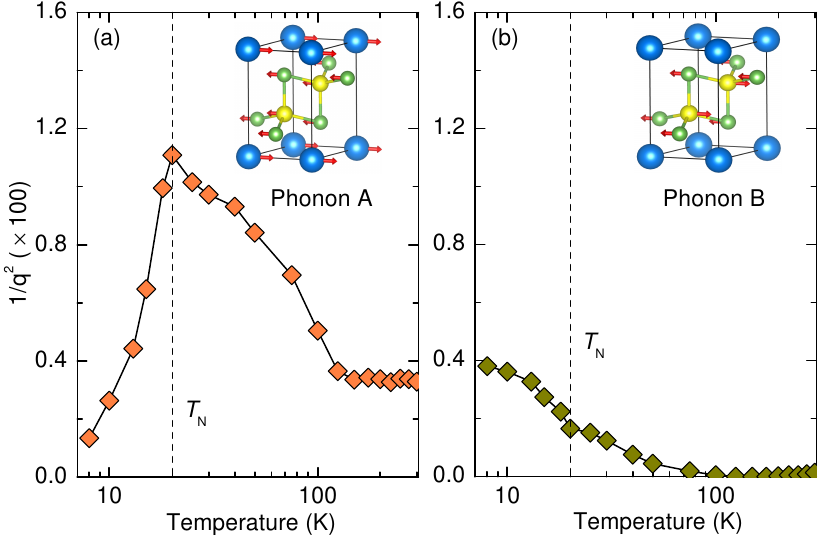}
\caption{(a) and (b) Temperature dependence of the Fano parameter $1/q^2$ for the infrared-active phonon modes at around 95 and 190 cm$^{-1}$, respectively. The insets show displacement patterns of the phonon modes.}
\label{fig4}
\end{figure}

Finally, we turn to the dynamics of the infrared-active phonons in EuZn$_2$As$_2$. As shown in Fig.~\ref{fig2}, both phonon modes exhibit asymmetric line shapes at low temperatures, which may result from coupling to either spins or charge excitations. To model these line shapes, the Fano oscillators are applied, and the obtained Fano parameters $1/q^2$ are plotted in Fig.~\ref{fig4}(a) for the 95\icm\ mode (phonon A) and Fig.~\ref{fig4}(b) for the 190\icm\ mode (phonon B). Notably, $1/q^2$ behaves differently across the AFM transition for these two modes. At high temperatures, $1/q^2$ for both modes is either small or zero, indicating weak or no coupling. As the temperature decreases, $1/q^2$ rises below 100 K, reflecting an enhancement in coupling strength. However, for phonon A, $1/q^2$ drops suddenly below $T_{\rm N} = 20$ K, while for phonon B, $1/q^2$ continues to rise without any anomaly. Examining the atomic displacements of the phonon modes, we find that phonon A involves the vibrations of Eu atoms, while phonon B is dominated by the in-plane vibrations of Zn and As atoms. As a result, the lattice vibrations of phonon A are strongly coupled to the fluctuations of Eu spins. Below $T_{\rm N}$, the spin-lattice coupling weakens due to the suppression of spin fluctuations. In contrast, phonon B, which does not involve Eu atoms, is unaffected by spin fluctuations. The asymmetric line shape of phonon B is primarily attributed to the coupling between phonon and charge excitations, which correlates with the enhancement of the $\alpha$ peak below 100 K. Moreover, as detailed in the Supplemental Materials~\footnotemark[1], other phonon parameters also show anomalies near $T_{\rm N}$, reflecting the complex interplay of spin, lattice, and charge excitations in this system.

To summarize, magnetic order often induces significant changes in the electronic structure of magnetic topological materials~\cite{otrokov2019prediction,liu2020robust,deng2020quantum,li2019intrinsic,wang2018large,yang2020magnetization}, typically through exchange coupling between local spins and the $spd$ electrons, leading to band splitting or shifts that can reduce or close the bulk band gap. In EuZn$_2$As$_2$, we observe a significant decrease in the band gap driven by the AFM order, but no nontrivial topological band inversion is detected. However, as discussed in the Supplemental Materials~\footnotemark[1], a topological transition can be induced in this system by tuning the Hubbard $U$. Interestingly, a recent optical study of EuCd$_2$As$_2$~\cite{santos2023eucd} showed that applying a magnetic field induced a remarkable 15\% reduction in the band gap. This suggests that a similar or even more pronounced change could occur in EuZn$_2$As$_2$ under an external magnetic field or high pressure~\cite{luo2023colossal}. Therefore, these external factors may further modulate the interplay between spin, charge, and lattice degrees of freedom, giving rise to novel and fascinating electronic properties and possible topological phase transitions in EuZn$_2$As$_2$.

%%%%%%%%%%%%%%%%%%%%%%%%%%%%%%%%%%%%%%%%%%%%%%%%%%%%%%%%%%%%%%%%%%%%%%%%%%%%%%
%
% Acknowledgment
%
\begin{acknowledgments}
This work was supported by the National Key Research and Development Program of China (Grants No. 2024YFA1408301, No. 2022YFA1403900, No. 2023YFA1406002 and No. 2023YFF0718400), the National Natural Science Foundation of China (Grant No. 12274442) and the Guangdong basic and applied basic research foundation (Grant No. 2023B151520013). P.M. and C.B. acknowledge funding by the Swiss National Science Foundation through Grants No. 200021-214905.
\end{acknowledgments}

%%%%%%%%%%%%%%%%%%%%%%%%%%%%%%%%%%%%%%%%%%%%%%%%%%%%%%%%%%%%%%%%%%%%%%%%%%%%%%%
% The bibliography (BibTeX)

%%%%%%%%%% Merge with supplemental materials %%%%%%%%%%
\pagebreak
\widetext
\begin{center}
\textbf{\large Supplemental Materials for
\\
Spectroscopic signatures of magnetization-induced band renormalization and strong spin-charge-lattice coupling in EuZn$_2$As$_2$}
\end{center}
%%%%%%%%%% Merge with supplemental materials %%%%%%%%%%
%%%%%%%%%% Prefix a "S" to all equations, figures, tables and reset the counter %%%%%%%%%%
\setcounter{equation}{0}
\setcounter{figure}{0}
\setcounter{table}{0}
\setcounter{page}{1}
\makeatletter
\renewcommand{\theequation}{S\arabic{equation}}
\renewcommand{\thefigure}{S\arabic{figure}}
\renewcommand{\bibnumfmt}[1]{[S#1]}
\renewcommand{\citenumfont}[1]{S#1}

\subsection{Sample synthesis and experimental details}
High-quality single crystals of EuZn$_2$As$_2$ were synthesized via the Sn-flux method~\cite{wang2022anisotropy,blawat2022unusual,bukowski2022canted}. The obtained crystals with sizes up to 5 mm $\times$ 3 mm $\times$ 1.3 mm have the form of platelets with shiny surfaces. Electrical resistivity and magnetic susceptibility measurements were performed by using a Quantum Design Physical Properties Measurement System (PPMS) and a Magnetic Properties Measurement System (MPMS), respectively.

The frequency-dependent reflectivity $R(\omega)$ at a near-normal angle of incidence have been measured on a flat $ab$ surface of the as-grown single crystal by using a Bruker VERTEX 70v Fourier transform infrared (FTIR) spectrometer. An \emph{in situ} gold evaporation technique~\cite{homes1993technique} was used to obtain the absolute reflectivity of the sample. Reflectivity data from 30 to 25\,000\icm\ were collected at various temperatures down to 10 K by using a commercial optical $^4$He cryostat. Transmission spectra from 500 to 15\,000\icm\ were measured by a FTIR spectrometer (Bruker VERTEX 80v) at various temperatures down to 10 K by using a commercial optical $^4$He cryostat, and the transmittance is obtained by the ratio to a reference spectrum taken in the absence of the sample. The room-temperature optical response function in the near-infrared to ultraviolet range (4000 to 50\,000 cm$^{-1}$) was measured with a commercial ellipsometer (Woollam VASE).

\subsection{Drude-Lorentz model and Kramers-Kronig analysis}
For a quantitative analysis of the optical data, we employed a Drude-Lorentz model to fit the measured low-frequency reflectivity $R(\omega)$. The model can be expressed in terms of the complex dielectric function as:
\begin{equation}
\label{Drude-Lorentz}
\Tilde{\epsilon}(\omega)=\epsilon_{\infty} + \frac{\omega_{p}^2}{\omega_{0}^{2}-\omega^{2}-i\gamma\omega}
+ \sum_{j}\left[\frac{\omega_{p,j}^2}{\omega_{0,j}^{2}-\omega^{2}-i\gamma_j\omega} \left( 1+i\frac{\omega_{q,j}}{\omega} \right)^2 + \left( \frac{\omega_{p,j}\omega_{q,j}}{\omega_{0,j} \omega} \right)^2\right].
\end{equation}
It describes the optical response of a set of Lorentz and Fano oscillators. Each Lorentz oscillator is characterized by the resonance frequency $\omega_{0}$, linewidth $\gamma$, and plasma frequency $\omega_{p}$. For the Drude response, $\omega_{0}$ is set to zero. The asymmetry in the Fano oscillator is controlled by a dimensionless factor $1/q=\omega_q/\omega_{0}$, where $\omega_q$ is the asymmetry-related frequency, which reflects the strength of the coupling between the phonon and the underlying optical electronic excitations. The parameter $\epsilon_{\infty}$ represents the high-frequency dielectric constant. The complex conductivity is related to the dielectric function by $\Tilde{\sigma}(\omega) = \sigma_1(\omega) + i\sigma_2(\omega) = -2\pi i\omega[\Tilde{\varepsilon}(\omega) - \varepsilon_\infty]/Z_0$ (in unists of $\Omega^{-1}\mathrm{cm}^{-1}$), where $Z_0 = 377~\Omega$ is the vacuum impedance. The normal-incidence reflectivity $R(\omega)$ is given by $R = \mid\frac{1-\sqrt{\Tilde{\varepsilon}}}{1+\sqrt{\Tilde{\varepsilon}}}\mid^{2}$.

\begin{figure}[htp]
\includegraphics[width=\textwidth]{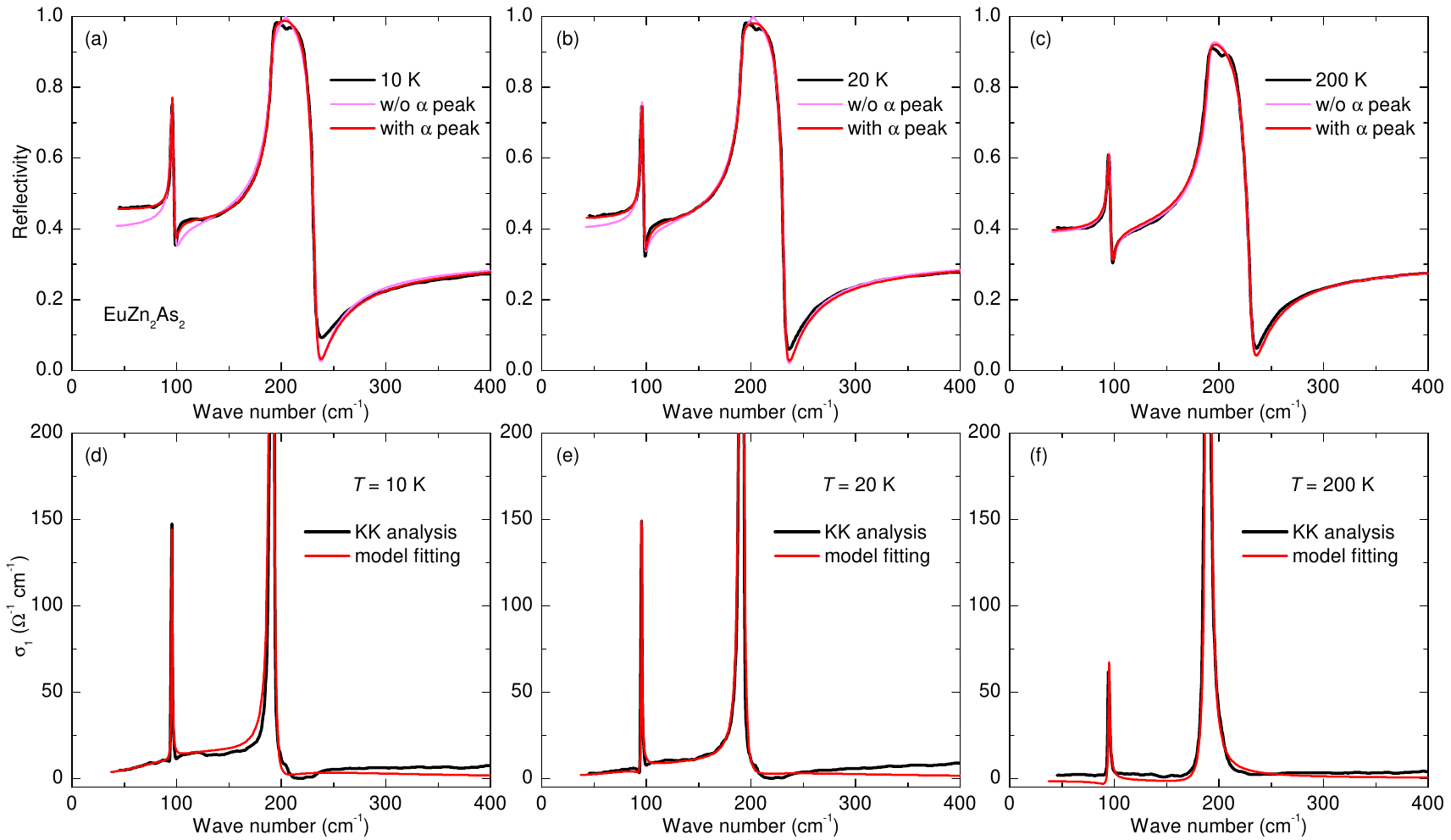}
\caption{(a)-(c) Drude-Lorentz fitting of the reflectivity at 10, 20, and 200 K, respectively. (d)-(f) Comparison of the $\sigma_1(\omega)$ obtained from model fitting and Kramers-Kronig analysis.}
\label{FigS1}
\end{figure}

Figures~\ref{FigS1}(a--c) present the fitting results of $R(\omega)$ at 10, 20, and 200 K, respectively. Firstly, two Fano oscillators are used to fit the two phonon modes (labeled as A and B), which reproduce the measured $R(\omega)$ well, except for the $\alpha$ peak feature and a lower reflectivity value in the zero-frequency limit. To improve the fit, an additional Lorentz oscillator was added to capture the $\alpha$ peak feature.

Alternatively, the optical conductivity $\sigma_1(\omega)$ can be derived from the Kramers-Kronig analysis of $R(\omega)$~\cite{dressel2002electrodynamics} with appropriate extrapolations at both low and high frequencies. Due to the insulating and transparent nature of EuZn$_2$As$_2$ in the mid-infrared region, the low-frequency extrapolation is set as a constant, and the extrapolation above 1\,000\icm\ is also set as a constant. The obtained far-infrared $\sigma_1(\omega)$ spectra are shown in Fig.~2(c) in the main text. We present the $\sigma_1(\omega)$ obtained from both methods in Fig.~\ref{FigS1}(d--f) at three representative temperatures 10, 20 and 200 K, respectively, where the model fitting $\sigma_1(\omega)$ agrees well with the data obtained via Kramers-Kronig analysis.

\begin{figure}[htp]
\includegraphics[width=\textwidth]{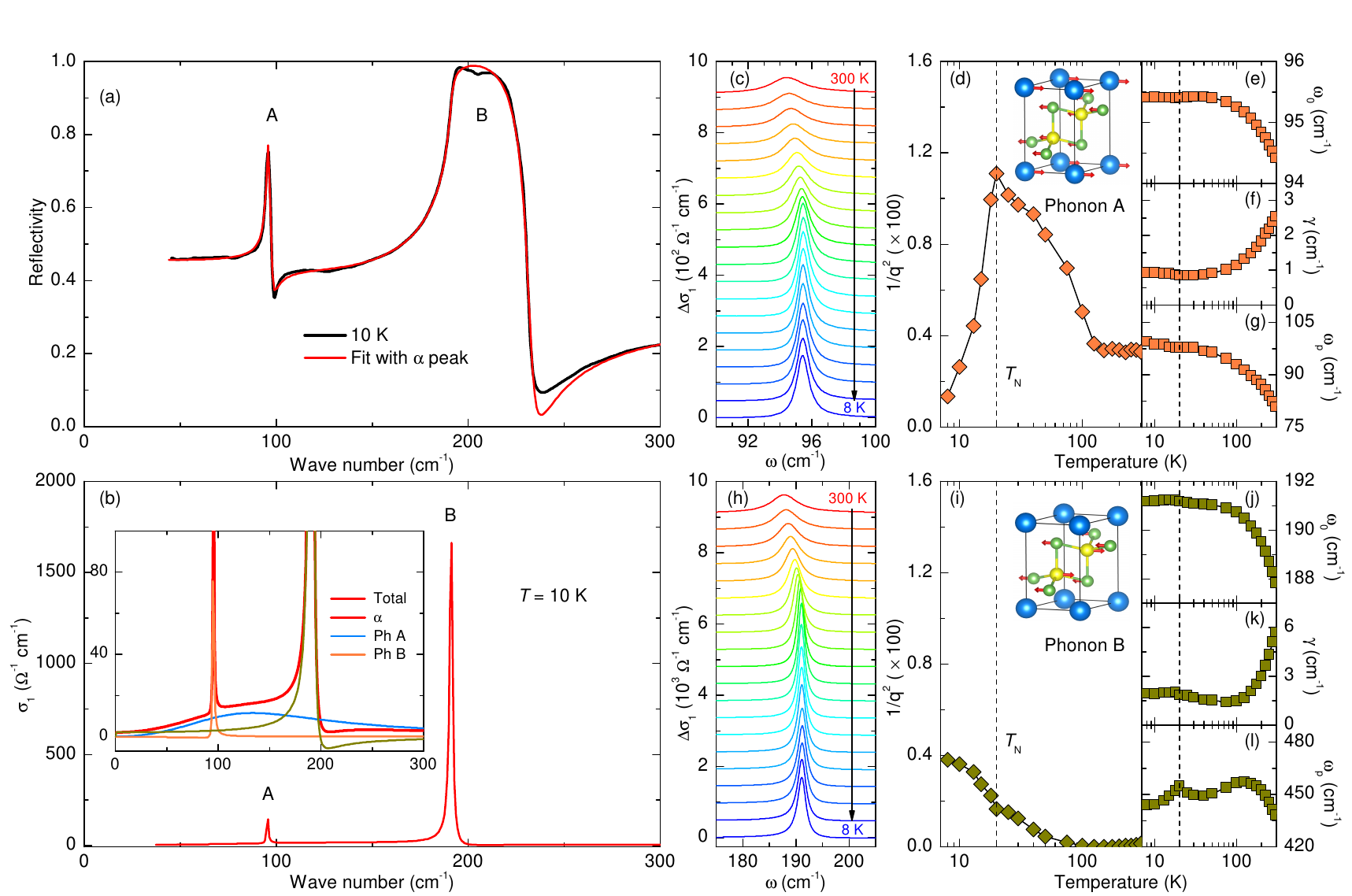}
\caption{(a) Drude-Lorentz fitting of the reflectivity at 10 K. (b) The corresponding $\sigma_1(\omega)$ obtained by Drude-Lorentz fitting. Inset shows the decomposition of the $\sigma_1(\omega)$. (c) and (h) Temperature-dependent line shape of the infrared-active phonon modes at around 95 and 190\icm, respectively. (d--g) Temperature dependence of the Fano parameter $1/q^2$, resonance frequency $\omega_0$, linewidth $\gamma$, and oscillator strength $\omega_p$ for phonon mode at around 95 cm$^{-1}$. The corresponding data for the mode around 190 cm$^{-1}$ are shown in (i--l). The insets show displacement patterns of the phonon modes.}
\label{FigS2}
\end{figure}

Figures~\ref{FigS2}(a) and \ref{FigS2}(b) show the Drude-Lorentz fitting of the reflectivity  and the corresponding $\sigma_1(\omega)$ at 10 K, which consists of two phonon modes with asymmetric line shape and the $\alpha$ peak centered around 130 cm$^{-1}$. The temperature dependences of the fitting parameters for the $\alpha$ peak are shown in Figs.~2(d--f) in the main text. Figures~\ref{FigS2}(c) and ~\ref{FigS2}(h) show the temperature dependence of the $\sigma_1(\omega)$ for Phonon A and B, respectively. The fitting parameters for phonon A are summarized in Figs.~\ref{FigS2}(d--g) and those for phonon B in Figs.~\ref{FigS2}(i--l).

\subsection{Electronic band calculations}
The first-principles calculations were carried out using the Vienna Ab-initio Simulation Package (VASP)~\cite{VASP_PW,VASP_ultrasoft}, employing a plane-wave basis set and projector augmented-wave method~\cite{VASP_PAW} to calculate the electronic band structure. The Perdew-Burke-Ernzerhof (PBE) exchange-correlation functional~\cite{VASP_PBE} within the generalized gradient approximation (GGA) was utilized. To account for the strong correlation effects of Eu-4$f$ electrons in the antiferromagnetic (AFM) phase, we employed the GGA+Hubbard-$U$ method~\cite{VASP_Hubbard_U}. We examined different $U$ values of 4, 5, and 6 eV, and found the best agreement with experiments when $U$ was set to 5 eV. Therefore, we employed the $U$ value of 5 eV to conduct the calculations of the AFM phase, which were compared with those of the paramagnetic (PM) phase. The plane-wave cutoff energy was set to 500 eV, and the convergence criterion for the total energy was set to 10$^{-7}$ eV. For both the PM and AFM phases, a 15$\times$15$\times$9 Monkhorst-Pack $k$-point mesh of the first Brillouin zone was employed for the integration. All results presented in this letter include the effect of spin-orbit coupling (SOC).

\subsection{Optical conductivity calculations}
Given the much smaller impact of indirect band gap transitions on optical conductivity in comparison to direct transitions, we focus solely on the contribution of direct transitions in the following calculation. In the calculation of optical conductivity, within the independent electron approximation, we take the minimal coupling Hamiltonian under the Coulomb gauge $H=\frac{1}{2 m}(p+q A)^2+V(r)$ and treat the electromagnetic field as a time-dependent perturbation. By employing Fermi's golden rule, we calculate the transition matrix elements from each occupied state to unoccupied state. Consequently, we derive the optical conductivity formula using the Kubo-Greenwood method~\cite{optical-kubo},

\begin{equation}
\begin{aligned}
\sigma_{\alpha \beta}(\hbar \omega)=
&\frac{i e^2 \hbar}{N_k \Omega_c} \sum_{\mathbf{k}} \sum_{n, m} \frac{f_{m \mathbf{k}}-f_{n \mathbf{k}}}{\varepsilon_{m \mathbf{k}}-\varepsilon_{n \mathbf{k}}} \\
&\times \frac{\left\langle\psi_{n \mathbf{k}}\left|v_\alpha\right| \psi_{m \mathbf{k}}\right\rangle\left\langle\psi_{m \mathbf{k}}\left|v_\beta\right| \psi_{n \mathbf{k}}\right\rangle}{\varepsilon_{m \mathbf{k}}-\varepsilon_{n \mathbf{k}}-(\hbar \omega+i \eta)},
\end{aligned}
\end{equation}
where $\alpha\beta$ are Cartesian coordinates, $n$ and $m$ represent the label of bands which are not equal, $\omega$ is the photon frequency, $N_k$ is the number of $k$-points, and $\Omega_c$ is the volume of the unit cell, $f_{\mathbf{k}}$ is the Fermi-Dirac distribution function, $\mathbf{v}$ is the velocity operator,
\begin{equation}
\left\langle\psi_{n\mathbf{k}}|\mathbf{v}|\psi_{m\mathbf{k}}\right\rangle=-\frac{i}{\hbar}(\varepsilon_{mk}-\varepsilon_{nk})\mathbf{A}_{nm}(\mathbf{k}),
\end{equation}
where $\mathbf{A}_{nm}(\mathbf{k})$ is the Berry connection, satisfying
\begin{equation}
    \mathbf{A}_{nm}(\mathbf{k})=~\left\langle u_{n\mathbf{k}}\left|i\mathbf{\nabla}_\mathbf{k}\right|u_{m\mathbf{k}}\right\rangle=\mathbf{A}_{mn}^\ast(\mathbf{k}).
\end{equation}

We use the \textit{postw90} module in \small{WANNIER90}~\cite{Wannier90} to implement the
optical conductivity calculation. Various energy smearing parameters are employed to achieve appropriate curves and a 150$\times$150$\times$150 $k$-mesh is used in all the optical conductivity calculations.

\subsection{Topological crystalline insulator driven by Hubbard $U$}
In crystal EuZn$_2$As$_2$, which belongs to space group $P\bar{3}m1$ (No. 164), the Eu atoms possess magnetic moments that significantly reduce the symmetry of the system, leading to a magnetic space group classified as No.~12.63 in the ground state AFM[100]. Using the magnetic indicator method as described in \cite{indicator_mag}, we computed the indicator for this magnetic space group, denoted as $z_{2 P}^{\prime}$:
\begin{equation}
z_{2 P}^{\prime}=\sum_{\mathbf{k} \in \mathrm{TRIM}} \frac{1}{4}\left(N_{\mathbf{k}}^{-}-N_{\mathbf{k}}^{+}\right) \bmod 2.
\end{equation}
Our results show that with a Hubbard $U$ of 6 eV, the system is trivial. When $U$ = 5 eV, the gap almost closed at $\Gamma$, which can be regarded as a critical point of topological phase transition. However, at a $U$ value of 4 eV, the system becomes a topological crystalline insulator (TCI) characterized by a $Z_2$ invariant. This indicates that by tuning the parameter $U$, we can drive a topological transition in the system. Specifically, when the correlation effect is relatively strong, the system transitions from a topological phase to a trivial phase.

%\bibliography{biblio}

\begin{thebibliography}{45}%
\makeatletter
\providecommand \@ifxundefined [1]{%
 \@ifx{#1\undefined}
}%
\providecommand \@ifnum [1]{%
 \ifnum #1\expandafter \@firstoftwo
 \else \expandafter \@secondoftwo
 \fi
}%
\providecommand \@ifx [1]{%
 \ifx #1\expandafter \@firstoftwo
 \else \expandafter \@secondoftwo
 \fi
}%
\providecommand \natexlab [1]{#1}%
\providecommand \enquote  [1]{``#1''}%
\providecommand \bibnamefont  [1]{#1}%
\providecommand \bibfnamefont [1]{#1}%
\providecommand \citenamefont [1]{#1}%
\providecommand \href@noop [0]{\@secondoftwo}%
\providecommand \href [0]{\begingroup \@sanitize@url \@href}%
\providecommand \@href[1]{\@@startlink{#1}\@@href}%
\providecommand \@@href[1]{\endgroup#1\@@endlink}%
\providecommand \@sanitize@url [0]{\catcode `\\12\catcode `\$12\catcode
  `\&12\catcode `\#12\catcode `\^12\catcode `\_12\catcode `\%12\relax}%
\providecommand \@@startlink[1]{}%
\providecommand \@@endlink[0]{}%
\providecommand \url  [0]{\begingroup\@sanitize@url \@url }%
\providecommand \@url [1]{\endgroup\@href {#1}{\urlprefix }}%
\providecommand \urlprefix  [0]{URL }%
\providecommand \Eprint [0]{\href }%
\providecommand \doibase [0]{https://doi.org/}%
\providecommand \selectlanguage [0]{\@gobble}%
\providecommand \bibinfo  [0]{\@secondoftwo}%
\providecommand \bibfield  [0]{\@secondoftwo}%
\providecommand \translation [1]{[#1]}%
\providecommand \BibitemOpen [0]{}%
\providecommand \bibitemStop [0]{}%
\providecommand \bibitemNoStop [0]{.\EOS\space}%
\providecommand \EOS [0]{\spacefactor3000\relax}%
\providecommand \BibitemShut  [1]{\csname bibitem#1\endcsname}%
\let\auto@bib@innerbib\@empty
%</preamble>
\bibitem [{\citenamefont {Yan}\ and\ \citenamefont
  {Felser}(2017)}]{yan2017topological}%
  \BibitemOpen
  \bibfield  {author} {\bibinfo {author} {\bibfnamefont {B.}~\bibnamefont
  {Yan}}\ and\ \bibinfo {author} {\bibfnamefont {C.}~\bibnamefont {Felser}},\
  }\bibfield  {title} {\bibinfo {title} {Topological materials: Weyl
  semimetals},\ }\href
  {https://doi.org/10.1146/annurev-conmatphys-031016-025458} {\bibfield
  {journal} {\bibinfo  {journal} {Annu. Rev. Condens. Matter Phys.}\ }\textbf
  {\bibinfo {volume} {8}},\ \bibinfo {pages} {337} (\bibinfo {year}
  {2017})}\BibitemShut {NoStop}%
\bibitem [{\citenamefont {Bernevig}\ \emph {et~al.}(2022)\citenamefont
  {Bernevig}, \citenamefont {Felser},\ and\ \citenamefont
  {Beidenkopf}}]{bernevig2022progress}%
  \BibitemOpen
  \bibfield  {author} {\bibinfo {author} {\bibfnamefont {B.~A.}\ \bibnamefont
  {Bernevig}}, \bibinfo {author} {\bibfnamefont {C.}~\bibnamefont {Felser}},\
  and\ \bibinfo {author} {\bibfnamefont {H.}~\bibnamefont {Beidenkopf}},\
  }\bibfield  {title} {\bibinfo {title} {{Progress and prospects in magnetic
  topological materials}},\ }\href {https://doi.org/10.1038/s41586-021-04105-x}
  {\bibfield  {journal} {\bibinfo  {journal} {Nature}\ }\textbf {\bibinfo
  {volume} {603}},\ \bibinfo {pages} {41} (\bibinfo {year} {2022})}\BibitemShut
  {NoStop}%
\bibitem [{\citenamefont {Wang}\ \emph {et~al.}(2023)\citenamefont {Wang},
  \citenamefont {Zhang}, \citenamefont {Zeng}, \citenamefont {Sun},
  \citenamefont {Hao}, \citenamefont {Cai}, \citenamefont {Rong}, \citenamefont
  {Zhang}, \citenamefont {Liu}, \citenamefont {Ma} \emph
  {et~al.}}]{wang2023intrinsic}%
  \BibitemOpen
  \bibfield  {author} {\bibinfo {author} {\bibfnamefont {Y.}~\bibnamefont
  {Wang}}, \bibinfo {author} {\bibfnamefont {F.}~\bibnamefont {Zhang}},
  \bibinfo {author} {\bibfnamefont {M.}~\bibnamefont {Zeng}}, \bibinfo {author}
  {\bibfnamefont {H.}~\bibnamefont {Sun}}, \bibinfo {author} {\bibfnamefont
  {Z.}~\bibnamefont {Hao}}, \bibinfo {author} {\bibfnamefont {Y.}~\bibnamefont
  {Cai}}, \bibinfo {author} {\bibfnamefont {H.}~\bibnamefont {Rong}}, \bibinfo
  {author} {\bibfnamefont {C.}~\bibnamefont {Zhang}}, \bibinfo {author}
  {\bibfnamefont {C.}~\bibnamefont {Liu}}, \bibinfo {author} {\bibfnamefont
  {X.}~\bibnamefont {Ma}}, \emph {et~al.},\ }\bibfield  {title} {\bibinfo
  {title} {{Intrinsic magnetic topological materials}},\ }\href
  {https://doi.org/10.1007/s11467-022-1250-6} {\bibfield  {journal} {\bibinfo
  {journal} {Front. Phys.}\ }\textbf {\bibinfo {volume} {18}},\ \bibinfo
  {pages} {21304} (\bibinfo {year} {2023})}\BibitemShut {NoStop}%
\bibitem [{\citenamefont {Chen}\ \emph {et~al.}(2024)\citenamefont {Chen},
  \citenamefont {Dong},\ and\ \citenamefont {Wang}}]{chen2024recent}%
  \BibitemOpen
  \bibfield  {author} {\bibinfo {author} {\bibfnamefont {X.}~\bibnamefont
  {Chen}}, \bibinfo {author} {\bibfnamefont {S.}~\bibnamefont {Dong}},\ and\
  \bibinfo {author} {\bibfnamefont {Z.-C.}\ \bibnamefont {Wang}},\ }\bibfield
  {title} {\bibinfo {title} {{Recent advances in understanding and manipulating
  magnetic and electronic properties of EuM$_2$X$_2$ (M=Zn, Cd; X=P, As)}},\
  }\href {https://doi.org/10.1088/1361-648X/ad882b} {\bibfield  {journal}
  {\bibinfo  {journal} {J. Phys.: Condens. Matter}\ }\textbf {\bibinfo {volume}
  {37}},\ \bibinfo {pages} {033001} (\bibinfo {year} {2024})}\BibitemShut
  {NoStop}%
\bibitem [{\citenamefont {Ma}\ \emph {et~al.}(2019)\citenamefont {Ma},
  \citenamefont {Nie}, \citenamefont {Yi}, \citenamefont {Jandke},
  \citenamefont {Shang}, \citenamefont {Yao}, \citenamefont {Naamneh},
  \citenamefont {Yan}, \citenamefont {Sun}, \citenamefont {Chikina} \emph
  {et~al.}}]{ma2019spin}%
  \BibitemOpen
  \bibfield  {author} {\bibinfo {author} {\bibfnamefont {J.-Z.}\ \bibnamefont
  {Ma}}, \bibinfo {author} {\bibfnamefont {S.}~\bibnamefont {Nie}}, \bibinfo
  {author} {\bibfnamefont {C.}~\bibnamefont {Yi}}, \bibinfo {author}
  {\bibfnamefont {J.}~\bibnamefont {Jandke}}, \bibinfo {author} {\bibfnamefont
  {T.}~\bibnamefont {Shang}}, \bibinfo {author} {\bibfnamefont {M.-Y.}\
  \bibnamefont {Yao}}, \bibinfo {author} {\bibfnamefont {M.}~\bibnamefont
  {Naamneh}}, \bibinfo {author} {\bibfnamefont {L.}~\bibnamefont {Yan}},
  \bibinfo {author} {\bibfnamefont {Y.}~\bibnamefont {Sun}}, \bibinfo {author}
  {\bibfnamefont {A.}~\bibnamefont {Chikina}}, \emph {et~al.},\ }\bibfield
  {title} {\bibinfo {title} {{Spin fluctuation induced Weyl semimetal state in
  the paramagnetic phase of EuCd$_2$As$_2$}},\ }\href
  {https://doi.org/10.1126/sciadv.aaw4718} {\bibfield  {journal} {\bibinfo
  {journal} {Sci. Adv.}\ }\textbf {\bibinfo {volume} {5}},\ \bibinfo {pages}
  {eaaw4718} (\bibinfo {year} {2019})}\BibitemShut {NoStop}%
\bibitem [{\citenamefont {Soh}\ \emph {et~al.}(2019)\citenamefont {Soh},
  \citenamefont {De~Juan}, \citenamefont {Vergniory}, \citenamefont
  {Schr{\"o}ter}, \citenamefont {Rahn}, \citenamefont {Yan}, \citenamefont
  {Jiang}, \citenamefont {Bristow}, \citenamefont {Reiss}, \citenamefont
  {Blandy} \emph {et~al.}}]{soh2019ideal}%
  \BibitemOpen
  \bibfield  {author} {\bibinfo {author} {\bibfnamefont {J.-R.}\ \bibnamefont
  {Soh}}, \bibinfo {author} {\bibfnamefont {F.}~\bibnamefont {De~Juan}},
  \bibinfo {author} {\bibfnamefont {M.}~\bibnamefont {Vergniory}}, \bibinfo
  {author} {\bibfnamefont {N.}~\bibnamefont {Schr{\"o}ter}}, \bibinfo {author}
  {\bibfnamefont {M.}~\bibnamefont {Rahn}}, \bibinfo {author} {\bibfnamefont
  {D.}~\bibnamefont {Yan}}, \bibinfo {author} {\bibfnamefont {J.}~\bibnamefont
  {Jiang}}, \bibinfo {author} {\bibfnamefont {M.}~\bibnamefont {Bristow}},
  \bibinfo {author} {\bibfnamefont {P.}~\bibnamefont {Reiss}}, \bibinfo
  {author} {\bibfnamefont {J.}~\bibnamefont {Blandy}}, \emph {et~al.},\
  }\bibfield  {title} {\bibinfo {title} {{Ideal Weyl semimetal induced by
  magnetic exchange}},\ }\href {https://doi.org/10.1103/PhysRevB.100.201102}
  {\bibfield  {journal} {\bibinfo  {journal} {Phys. Rev. B}\ }\textbf {\bibinfo
  {volume} {100}},\ \bibinfo {pages} {201102} (\bibinfo {year}
  {2019})}\BibitemShut {NoStop}%
\bibitem [{\citenamefont {Wang}\ \emph {et~al.}(2019)\citenamefont {Wang},
  \citenamefont {Jo}, \citenamefont {Kuthanazhi}, \citenamefont {Wu},
  \citenamefont {McQueeney}, \citenamefont {Kaminski},\ and\ \citenamefont
  {Canfield}}]{wang2019single}%
  \BibitemOpen
  \bibfield  {author} {\bibinfo {author} {\bibfnamefont {L.-L.}\ \bibnamefont
  {Wang}}, \bibinfo {author} {\bibfnamefont {N.~H.}\ \bibnamefont {Jo}},
  \bibinfo {author} {\bibfnamefont {B.}~\bibnamefont {Kuthanazhi}}, \bibinfo
  {author} {\bibfnamefont {Y.}~\bibnamefont {Wu}}, \bibinfo {author}
  {\bibfnamefont {R.~J.}\ \bibnamefont {McQueeney}}, \bibinfo {author}
  {\bibfnamefont {A.}~\bibnamefont {Kaminski}},\ and\ \bibinfo {author}
  {\bibfnamefont {P.~C.}\ \bibnamefont {Canfield}},\ }\bibfield  {title}
  {\bibinfo {title} {{Single pair of Weyl fermions in the half-metallic
  semimetal EuCd$_2$As$_2$}},\ }\href
  {https://doi.org/10.1103/PhysRevB.99.245147} {\bibfield  {journal} {\bibinfo
  {journal} {Phys. Rev. B}\ }\textbf {\bibinfo {volume} {99}},\ \bibinfo
  {pages} {245147} (\bibinfo {year} {2019})}\BibitemShut {NoStop}%
\bibitem [{\citenamefont {Ma}\ \emph {et~al.}(2020)\citenamefont {Ma},
  \citenamefont {Wang}, \citenamefont {Nie}, \citenamefont {Yi}, \citenamefont
  {Xu}, \citenamefont {Li}, \citenamefont {Jandke}, \citenamefont {Wulfhekel},
  \citenamefont {Huang}, \citenamefont {West} \emph
  {et~al.}}]{ma2020emergence}%
  \BibitemOpen
  \bibfield  {author} {\bibinfo {author} {\bibfnamefont {J.}~\bibnamefont
  {Ma}}, \bibinfo {author} {\bibfnamefont {H.}~\bibnamefont {Wang}}, \bibinfo
  {author} {\bibfnamefont {S.}~\bibnamefont {Nie}}, \bibinfo {author}
  {\bibfnamefont {C.}~\bibnamefont {Yi}}, \bibinfo {author} {\bibfnamefont
  {Y.}~\bibnamefont {Xu}}, \bibinfo {author} {\bibfnamefont {H.}~\bibnamefont
  {Li}}, \bibinfo {author} {\bibfnamefont {J.}~\bibnamefont {Jandke}}, \bibinfo
  {author} {\bibfnamefont {W.}~\bibnamefont {Wulfhekel}}, \bibinfo {author}
  {\bibfnamefont {Y.}~\bibnamefont {Huang}}, \bibinfo {author} {\bibfnamefont
  {D.}~\bibnamefont {West}}, \emph {et~al.},\ }\bibfield  {title} {\bibinfo
  {title} {{Emergence of Nontrivial Low-Energy Dirac Fermions in
  Antiferromagnetic EuCd$_2$As$_2$}},\ }\href
  {https://doi.org/10.1002/adma.201907565} {\bibfield  {journal} {\bibinfo
  {journal} {Adv. Mater.}\ }\textbf {\bibinfo {volume} {32}},\ \bibinfo {pages}
  {1907565} (\bibinfo {year} {2020})}\BibitemShut {NoStop}%
\bibitem [{\citenamefont {Taddei}\ \emph {et~al.}(2022)\citenamefont {Taddei},
  \citenamefont {Yin}, \citenamefont {Sanjeewa}, \citenamefont {Li},
  \citenamefont {Xing}, \citenamefont {Dela~Cruz}, \citenamefont {Phelan},
  \citenamefont {Sefat},\ and\ \citenamefont {Parker}}]{taddei2022single}%
  \BibitemOpen
  \bibfield  {author} {\bibinfo {author} {\bibfnamefont {K.}~\bibnamefont
  {Taddei}}, \bibinfo {author} {\bibfnamefont {L.}~\bibnamefont {Yin}},
  \bibinfo {author} {\bibfnamefont {L.}~\bibnamefont {Sanjeewa}}, \bibinfo
  {author} {\bibfnamefont {Y.}~\bibnamefont {Li}}, \bibinfo {author}
  {\bibfnamefont {J.}~\bibnamefont {Xing}}, \bibinfo {author} {\bibfnamefont
  {C.}~\bibnamefont {Dela~Cruz}}, \bibinfo {author} {\bibfnamefont
  {D.}~\bibnamefont {Phelan}}, \bibinfo {author} {\bibfnamefont
  {A.}~\bibnamefont {Sefat}},\ and\ \bibinfo {author} {\bibfnamefont
  {D.}~\bibnamefont {Parker}},\ }\bibfield  {title} {\bibinfo {title} {{Single
  pair of Weyl nodes in the spin-canted structure of EuCd$_2$As$_2$}},\ }\href
  {https://doi.org/10.1103/PhysRevB.105.L140401} {\bibfield  {journal}
  {\bibinfo  {journal} {Phys. Rev. B}\ }\textbf {\bibinfo {volume} {105}},\
  \bibinfo {pages} {L140401} (\bibinfo {year} {2022})}\BibitemShut {NoStop}%
\bibitem [{\citenamefont {Xu}\ \emph {et~al.}(2019)\citenamefont {Xu},
  \citenamefont {Song}, \citenamefont {Wang}, \citenamefont {Weng},\ and\
  \citenamefont {Dai}}]{xu2019higher}%
  \BibitemOpen
  \bibfield  {author} {\bibinfo {author} {\bibfnamefont {Y.}~\bibnamefont
  {Xu}}, \bibinfo {author} {\bibfnamefont {Z.}~\bibnamefont {Song}}, \bibinfo
  {author} {\bibfnamefont {Z.}~\bibnamefont {Wang}}, \bibinfo {author}
  {\bibfnamefont {H.}~\bibnamefont {Weng}},\ and\ \bibinfo {author}
  {\bibfnamefont {X.}~\bibnamefont {Dai}},\ }\bibfield  {title} {\bibinfo
  {title} {{Higher-order topology of the axion insulator EuIn$_2$As$_2$}},\
  }\href {https://doi.org/10.1103/PhysRevLett.122.256402} {\bibfield  {journal}
  {\bibinfo  {journal} {Phys. Rev. Lett.}\ }\textbf {\bibinfo {volume} {122}},\
  \bibinfo {pages} {256402} (\bibinfo {year} {2019})}\BibitemShut {NoStop}%
\bibitem [{\citenamefont {Li}\ \emph {et~al.}(2019{\natexlab{a}})\citenamefont
  {Li}, \citenamefont {Gao}, \citenamefont {Duan}, \citenamefont {Xu},
  \citenamefont {Zhu}, \citenamefont {Tian}, \citenamefont {Gao}, \citenamefont
  {Fan}, \citenamefont {Rao}, \citenamefont {Huang} \emph
  {et~al.}}]{li2019dirac}%
  \BibitemOpen
  \bibfield  {author} {\bibinfo {author} {\bibfnamefont {H.}~\bibnamefont
  {Li}}, \bibinfo {author} {\bibfnamefont {S.-Y.}\ \bibnamefont {Gao}},
  \bibinfo {author} {\bibfnamefont {S.-F.}\ \bibnamefont {Duan}}, \bibinfo
  {author} {\bibfnamefont {Y.-F.}\ \bibnamefont {Xu}}, \bibinfo {author}
  {\bibfnamefont {K.-J.}\ \bibnamefont {Zhu}}, \bibinfo {author} {\bibfnamefont
  {S.-J.}\ \bibnamefont {Tian}}, \bibinfo {author} {\bibfnamefont {J.-C.}\
  \bibnamefont {Gao}}, \bibinfo {author} {\bibfnamefont {W.-H.}\ \bibnamefont
  {Fan}}, \bibinfo {author} {\bibfnamefont {Z.-C.}\ \bibnamefont {Rao}},
  \bibinfo {author} {\bibfnamefont {J.-R.}\ \bibnamefont {Huang}}, \emph
  {et~al.},\ }\bibfield  {title} {\bibinfo {title} {{Dirac surface states in
  intrinsic magnetic topological insulators EuSn$_2$As$_2$ and
  MnBi$_{2n}$Te$_{3 n+ 1}$ }},\ }\href
  {https://doi.org/10.1103/PhysRevX.9.041039} {\bibfield  {journal} {\bibinfo
  {journal} {Phys. Rev. X}\ }\textbf {\bibinfo {volume} {9}},\ \bibinfo {pages}
  {041039} (\bibinfo {year} {2019}{\natexlab{a}})}\BibitemShut {NoStop}%
\bibitem [{\citenamefont {Jiang}\ and\ \citenamefont
  {Kauzlarich}(2006)}]{jiang2006colossal}%
  \BibitemOpen
  \bibfield  {author} {\bibinfo {author} {\bibfnamefont {J.}~\bibnamefont
  {Jiang}}\ and\ \bibinfo {author} {\bibfnamefont {S.~M.}\ \bibnamefont
  {Kauzlarich}},\ }\bibfield  {title} {\bibinfo {title} {{Colossal
  magnetoresistance in a rare earth Zintl compound with a new structure type:
  EuIn$_2$P$_2$}},\ }\href {https://doi.org/10.1021/cm0520362} {\bibfield
  {journal} {\bibinfo  {journal} {Chem. Mater.}\ }\textbf {\bibinfo {volume}
  {18}},\ \bibinfo {pages} {435} (\bibinfo {year} {2006})}\BibitemShut
  {NoStop}%
\bibitem [{\citenamefont {Zhang}\ \emph {et~al.}(2020)\citenamefont {Zhang},
  \citenamefont {Deng}, \citenamefont {Zhang}, \citenamefont {Wang},
  \citenamefont {Wang}, \citenamefont {Liu}, \citenamefont {Mei}, \citenamefont
  {Kumar}, \citenamefont {Schwier}, \citenamefont {Shimada}, \citenamefont
  {Chen},\ and\ \citenamefont {Shen}}]{Zhang2020PRB}%
  \BibitemOpen
  \bibfield  {author} {\bibinfo {author} {\bibfnamefont {Y.}~\bibnamefont
  {Zhang}}, \bibinfo {author} {\bibfnamefont {K.}~\bibnamefont {Deng}},
  \bibinfo {author} {\bibfnamefont {X.}~\bibnamefont {Zhang}}, \bibinfo
  {author} {\bibfnamefont {M.}~\bibnamefont {Wang}}, \bibinfo {author}
  {\bibfnamefont {Y.}~\bibnamefont {Wang}}, \bibinfo {author} {\bibfnamefont
  {C.}~\bibnamefont {Liu}}, \bibinfo {author} {\bibfnamefont {J.-W.}\
  \bibnamefont {Mei}}, \bibinfo {author} {\bibfnamefont {S.}~\bibnamefont
  {Kumar}}, \bibinfo {author} {\bibfnamefont {E.~F.}\ \bibnamefont {Schwier}},
  \bibinfo {author} {\bibfnamefont {K.}~\bibnamefont {Shimada}}, \bibinfo
  {author} {\bibfnamefont {C.}~\bibnamefont {Chen}},\ and\ \bibinfo {author}
  {\bibfnamefont {B.}~\bibnamefont {Shen}},\ }\bibfield  {title} {\bibinfo
  {title} {{In-plane antiferromagnetic moments and magnetic polaron in the
  axion topological insulator candidate
  ${\mathrm{EuIn}}_{2}{\mathrm{As}}_{2}$}},\ }\href
  {https://doi.org/10.1103/PhysRevB.101.205126} {\bibfield  {journal} {\bibinfo
   {journal} {Phys. Rev. B}\ }\textbf {\bibinfo {volume} {101}},\ \bibinfo
  {pages} {205126} (\bibinfo {year} {2020})}\BibitemShut {NoStop}%
\bibitem [{\citenamefont {Wang}\ \emph {et~al.}(2021)\citenamefont {Wang},
  \citenamefont {Rogers}, \citenamefont {Yao}, \citenamefont {Nichols},
  \citenamefont {Atay}, \citenamefont {Xu}, \citenamefont {Franklin},
  \citenamefont {Sochnikov}, \citenamefont {Ryan}, \citenamefont {Haskel} \emph
  {et~al.}}]{wang2021colossal}%
  \BibitemOpen
  \bibfield  {author} {\bibinfo {author} {\bibfnamefont {Z.-C.}\ \bibnamefont
  {Wang}}, \bibinfo {author} {\bibfnamefont {J.~D.}\ \bibnamefont {Rogers}},
  \bibinfo {author} {\bibfnamefont {X.}~\bibnamefont {Yao}}, \bibinfo {author}
  {\bibfnamefont {R.}~\bibnamefont {Nichols}}, \bibinfo {author} {\bibfnamefont
  {K.}~\bibnamefont {Atay}}, \bibinfo {author} {\bibfnamefont {B.}~\bibnamefont
  {Xu}}, \bibinfo {author} {\bibfnamefont {J.}~\bibnamefont {Franklin}},
  \bibinfo {author} {\bibfnamefont {I.}~\bibnamefont {Sochnikov}}, \bibinfo
  {author} {\bibfnamefont {P.~J.}\ \bibnamefont {Ryan}}, \bibinfo {author}
  {\bibfnamefont {D.}~\bibnamefont {Haskel}}, \emph {et~al.},\ }\bibfield
  {title} {\bibinfo {title} {Colossal magnetoresistance without mixed valence
  in a layered phosphide crystal},\ }\href
  {https://doi.org/10.1002/adma.202005755} {\bibfield  {journal} {\bibinfo
  {journal} {Adv. Mater.}\ }\textbf {\bibinfo {volume} {33}},\ \bibinfo {pages}
  {2005755} (\bibinfo {year} {2021})}\BibitemShut {NoStop}%
\bibitem [{\citenamefont {Yan}\ \emph {et~al.}(2022)\citenamefont {Yan},
  \citenamefont {Jiang}, \citenamefont {Xiao}, \citenamefont {Lu},
  \citenamefont {Song}, \citenamefont {Zhu}, \citenamefont {Luo}, \citenamefont
  {Sun},\ and\ \citenamefont {Yamashita}}]{yan2022field}%
  \BibitemOpen
  \bibfield  {author} {\bibinfo {author} {\bibfnamefont {J.}~\bibnamefont
  {Yan}}, \bibinfo {author} {\bibfnamefont {Z.~Z.}\ \bibnamefont {Jiang}},
  \bibinfo {author} {\bibfnamefont {R.~C.}\ \bibnamefont {Xiao}}, \bibinfo
  {author} {\bibfnamefont {W.}~\bibnamefont {Lu}}, \bibinfo {author}
  {\bibfnamefont {W.}~\bibnamefont {Song}}, \bibinfo {author} {\bibfnamefont
  {X.}~\bibnamefont {Zhu}}, \bibinfo {author} {\bibfnamefont {X.}~\bibnamefont
  {Luo}}, \bibinfo {author} {\bibfnamefont {Y.}~\bibnamefont {Sun}},\ and\
  \bibinfo {author} {\bibfnamefont {M.}~\bibnamefont {Yamashita}},\ }\bibfield
  {title} {\bibinfo {title} {{Field-induced topological Hall effect in
  antiferromagnetic axion insulator candidate EuIn$_2$As$_2$}},\ }\href
  {https://doi.org/10.1103/PhysRevResearch.4.013163} {\bibfield  {journal}
  {\bibinfo  {journal} {Phys. Rev. Res.}\ }\textbf {\bibinfo {volume} {4}},\
  \bibinfo {pages} {013163} (\bibinfo {year} {2022})}\BibitemShut {NoStop}%
\bibitem [{\citenamefont {Li}\ \emph {et~al.}(2021)\citenamefont {Li},
  \citenamefont {Gao}, \citenamefont {Chen}, \citenamefont {Chu}, \citenamefont
  {Nie}, \citenamefont {Ma}, \citenamefont {Han}, \citenamefont {Wu},
  \citenamefont {Li}, \citenamefont {Niu} \emph {et~al.}}]{li2021magnetic}%
  \BibitemOpen
  \bibfield  {author} {\bibinfo {author} {\bibfnamefont {H.}~\bibnamefont
  {Li}}, \bibinfo {author} {\bibfnamefont {W.}~\bibnamefont {Gao}}, \bibinfo
  {author} {\bibfnamefont {Z.}~\bibnamefont {Chen}}, \bibinfo {author}
  {\bibfnamefont {W.}~\bibnamefont {Chu}}, \bibinfo {author} {\bibfnamefont
  {Y.}~\bibnamefont {Nie}}, \bibinfo {author} {\bibfnamefont {S.}~\bibnamefont
  {Ma}}, \bibinfo {author} {\bibfnamefont {Y.}~\bibnamefont {Han}}, \bibinfo
  {author} {\bibfnamefont {M.}~\bibnamefont {Wu}}, \bibinfo {author}
  {\bibfnamefont {T.}~\bibnamefont {Li}}, \bibinfo {author} {\bibfnamefont
  {Q.}~\bibnamefont {Niu}}, \emph {et~al.},\ }\bibfield  {title} {\bibinfo
  {title} {{Magnetic properties of the layered magnetic topological insulator
  EuSn$_2$As$_2$}},\ }\href {https://doi.org/10.1103/PhysRevB.104.054435}
  {\bibfield  {journal} {\bibinfo  {journal} {Phys. Rev. B}\ }\textbf {\bibinfo
  {volume} {104}},\ \bibinfo {pages} {054435} (\bibinfo {year}
  {2021})}\BibitemShut {NoStop}%
\bibitem [{\citenamefont {Du}\ \emph {et~al.}(2022)\citenamefont {Du},
  \citenamefont {Yang}, \citenamefont {Nie}, \citenamefont {Wu}, \citenamefont
  {Li}, \citenamefont {Luo}, \citenamefont {Chen}, \citenamefont {Su},
  \citenamefont {Smidman}, \citenamefont {Shi} \emph
  {et~al.}}]{du2022consecutive}%
  \BibitemOpen
  \bibfield  {author} {\bibinfo {author} {\bibfnamefont {F.}~\bibnamefont
  {Du}}, \bibinfo {author} {\bibfnamefont {L.}~\bibnamefont {Yang}}, \bibinfo
  {author} {\bibfnamefont {Z.}~\bibnamefont {Nie}}, \bibinfo {author}
  {\bibfnamefont {N.}~\bibnamefont {Wu}}, \bibinfo {author} {\bibfnamefont
  {Y.}~\bibnamefont {Li}}, \bibinfo {author} {\bibfnamefont {S.}~\bibnamefont
  {Luo}}, \bibinfo {author} {\bibfnamefont {Y.}~\bibnamefont {Chen}}, \bibinfo
  {author} {\bibfnamefont {D.}~\bibnamefont {Su}}, \bibinfo {author}
  {\bibfnamefont {M.}~\bibnamefont {Smidman}}, \bibinfo {author} {\bibfnamefont
  {Y.}~\bibnamefont {Shi}}, \emph {et~al.},\ }\bibfield  {title} {\bibinfo
  {title} {Consecutive topological phase transitions and colossal
  magnetoresistance in a magnetic topological semimetal},\ }\href
  {https://doi.org/10.1038/s41535-022-00468-0} {\bibfield  {journal} {\bibinfo
  {journal} {npj Quantum Mater.}\ }\textbf {\bibinfo {volume} {7}},\ \bibinfo
  {pages} {65} (\bibinfo {year} {2022})}\BibitemShut {NoStop}%
\bibitem [{\citenamefont {Zhang}\ \emph {et~al.}(2023)\citenamefont {Zhang},
  \citenamefont {Du}, \citenamefont {Zheng}, \citenamefont {Luo}, \citenamefont
  {Wu}, \citenamefont {Zheng}, \citenamefont {Cui}, \citenamefont {Sun},
  \citenamefont {Liu}, \citenamefont {Shen} \emph
  {et~al.}}]{zhang2023electronic}%
  \BibitemOpen
  \bibfield  {author} {\bibinfo {author} {\bibfnamefont {H.}~\bibnamefont
  {Zhang}}, \bibinfo {author} {\bibfnamefont {F.}~\bibnamefont {Du}}, \bibinfo
  {author} {\bibfnamefont {X.}~\bibnamefont {Zheng}}, \bibinfo {author}
  {\bibfnamefont {S.}~\bibnamefont {Luo}}, \bibinfo {author} {\bibfnamefont
  {Y.}~\bibnamefont {Wu}}, \bibinfo {author} {\bibfnamefont {H.}~\bibnamefont
  {Zheng}}, \bibinfo {author} {\bibfnamefont {S.}~\bibnamefont {Cui}}, \bibinfo
  {author} {\bibfnamefont {Z.}~\bibnamefont {Sun}}, \bibinfo {author}
  {\bibfnamefont {Z.}~\bibnamefont {Liu}}, \bibinfo {author} {\bibfnamefont
  {D.}~\bibnamefont {Shen}}, \emph {et~al.},\ }\bibfield  {title} {\bibinfo
  {title} {{Electronic band reconstruction across the insulator-metal
  transition in colossally magnetoresistive EuCd$_2$P$_2$}},\ }\href
  {https://doi.org/10.1103/PhysRevB.108.L241115} {\bibfield  {journal}
  {\bibinfo  {journal} {Phys. Rev. B}\ }\textbf {\bibinfo {volume} {108}},\
  \bibinfo {pages} {L241115} (\bibinfo {year} {2023})}\BibitemShut {NoStop}%
\bibitem [{\citenamefont {Krebber}\ \emph {et~al.}(2023)\citenamefont
  {Krebber}, \citenamefont {Kopp}, \citenamefont {Garg}, \citenamefont
  {Kummer}, \citenamefont {Sichelschmidt}, \citenamefont {Schulz},
  \citenamefont {Poelchen}, \citenamefont {Mende}, \citenamefont {Virovets},
  \citenamefont {Warawa} \emph {et~al.}}]{krebber2023colossal}%
  \BibitemOpen
  \bibfield  {author} {\bibinfo {author} {\bibfnamefont {S.}~\bibnamefont
  {Krebber}}, \bibinfo {author} {\bibfnamefont {M.}~\bibnamefont {Kopp}},
  \bibinfo {author} {\bibfnamefont {C.}~\bibnamefont {Garg}}, \bibinfo {author}
  {\bibfnamefont {K.}~\bibnamefont {Kummer}}, \bibinfo {author} {\bibfnamefont
  {J.}~\bibnamefont {Sichelschmidt}}, \bibinfo {author} {\bibfnamefont
  {S.}~\bibnamefont {Schulz}}, \bibinfo {author} {\bibfnamefont
  {G.}~\bibnamefont {Poelchen}}, \bibinfo {author} {\bibfnamefont
  {M.}~\bibnamefont {Mende}}, \bibinfo {author} {\bibfnamefont {A.~V.}\
  \bibnamefont {Virovets}}, \bibinfo {author} {\bibfnamefont {K.}~\bibnamefont
  {Warawa}}, \emph {et~al.},\ }\bibfield  {title} {\bibinfo {title} {{Colossal
  magnetoresistance in EuZn$_2$P$_2$ and its electronic and magnetic
  structure}},\ }\href {https://doi.org/10.1103/PhysRevB.108.045116} {\bibfield
   {journal} {\bibinfo  {journal} {Phys. Rev. B}\ }\textbf {\bibinfo {volume}
  {108}},\ \bibinfo {pages} {045116} (\bibinfo {year} {2023})}\BibitemShut
  {NoStop}%
\bibitem [{\citenamefont {Wang}\ \emph {et~al.}(2022)\citenamefont {Wang},
  \citenamefont {Been}, \citenamefont {Gaudet}, \citenamefont {Alqasseri},
  \citenamefont {Fruhling}, \citenamefont {Yao}, \citenamefont {Stuhr},
  \citenamefont {Zhu}, \citenamefont {Ren}, \citenamefont {Cui} \emph
  {et~al.}}]{wang2022anisotropy}%
  \BibitemOpen
  \bibfield  {author} {\bibinfo {author} {\bibfnamefont {Z.-C.}\ \bibnamefont
  {Wang}}, \bibinfo {author} {\bibfnamefont {E.}~\bibnamefont {Been}}, \bibinfo
  {author} {\bibfnamefont {J.}~\bibnamefont {Gaudet}}, \bibinfo {author}
  {\bibfnamefont {G.~M.~A.}\ \bibnamefont {Alqasseri}}, \bibinfo {author}
  {\bibfnamefont {K.}~\bibnamefont {Fruhling}}, \bibinfo {author}
  {\bibfnamefont {X.}~\bibnamefont {Yao}}, \bibinfo {author} {\bibfnamefont
  {U.}~\bibnamefont {Stuhr}}, \bibinfo {author} {\bibfnamefont
  {Q.}~\bibnamefont {Zhu}}, \bibinfo {author} {\bibfnamefont {Z.}~\bibnamefont
  {Ren}}, \bibinfo {author} {\bibfnamefont {Y.}~\bibnamefont {Cui}}, \emph
  {et~al.},\ }\bibfield  {title} {\bibinfo {title} {{Anisotropy of the magnetic
  and transport properties of EuZn$_2$As$_2$}},\ }\href
  {https://doi.org/10.1103/PhysRevB.105.165122} {\bibfield  {journal} {\bibinfo
   {journal} {Phys. Rev. B}\ }\textbf {\bibinfo {volume} {105}},\ \bibinfo
  {pages} {165122} (\bibinfo {year} {2022})}\BibitemShut {NoStop}%
\bibitem [{\citenamefont {Blawat}\ \emph {et~al.}(2022)\citenamefont {Blawat},
  \citenamefont {Marshall}, \citenamefont {Singleton}, \citenamefont {Feng},
  \citenamefont {Cao}, \citenamefont {Xie},\ and\ \citenamefont
  {Jin}}]{blawat2022unusual}%
  \BibitemOpen
  \bibfield  {author} {\bibinfo {author} {\bibfnamefont {J.}~\bibnamefont
  {Blawat}}, \bibinfo {author} {\bibfnamefont {M.}~\bibnamefont {Marshall}},
  \bibinfo {author} {\bibfnamefont {J.}~\bibnamefont {Singleton}}, \bibinfo
  {author} {\bibfnamefont {E.}~\bibnamefont {Feng}}, \bibinfo {author}
  {\bibfnamefont {H.}~\bibnamefont {Cao}}, \bibinfo {author} {\bibfnamefont
  {W.}~\bibnamefont {Xie}},\ and\ \bibinfo {author} {\bibfnamefont
  {R.}~\bibnamefont {Jin}},\ }\bibfield  {title} {\bibinfo {title} {{Unusual
  electrical and magnetic properties in layered EuZn$_2$As$_2$}},\ }\href
  {https://doi.org/10.1002/qute.202200012} {\bibfield  {journal} {\bibinfo
  {journal} {Adv. Quantum Technol.}\ }\textbf {\bibinfo {volume} {5}},\
  \bibinfo {pages} {2200012} (\bibinfo {year} {2022})}\BibitemShut {NoStop}%
\bibitem [{\citenamefont {Bukowski}\ \emph {et~al.}(2022)\citenamefont
  {Bukowski}, \citenamefont {Rybicki}, \citenamefont {Babij}, \citenamefont
  {Przewo{\'z}nik}, \citenamefont {Gondek}, \citenamefont {{\.Z}ukrowski},\
  and\ \citenamefont {Kapusta}}]{bukowski2022canted}%
  \BibitemOpen
  \bibfield  {author} {\bibinfo {author} {\bibfnamefont {Z.}~\bibnamefont
  {Bukowski}}, \bibinfo {author} {\bibfnamefont {D.}~\bibnamefont {Rybicki}},
  \bibinfo {author} {\bibfnamefont {M.}~\bibnamefont {Babij}}, \bibinfo
  {author} {\bibfnamefont {J.}~\bibnamefont {Przewo{\'z}nik}}, \bibinfo
  {author} {\bibfnamefont {{\L}.}~\bibnamefont {Gondek}}, \bibinfo {author}
  {\bibfnamefont {J.}~\bibnamefont {{\.Z}ukrowski}},\ and\ \bibinfo {author}
  {\bibfnamefont {C.}~\bibnamefont {Kapusta}},\ }\bibfield  {title} {\bibinfo
  {title} {{Canted antiferromagnetic order in EuZn$_2$As$_2$ single
  crystals}},\ }\href {https://doi.org/10.1038/s41598-022-19026-6} {\bibfield
  {journal} {\bibinfo  {journal} {Sci. Rep.}\ }\textbf {\bibinfo {volume}
  {12}},\ \bibinfo {pages} {14718} (\bibinfo {year} {2022})}\BibitemShut
  {NoStop}%
\bibitem [{\citenamefont {Yi}\ \emph {et~al.}(2023)\citenamefont {Yi},
  \citenamefont {Zheng}, \citenamefont {Pan}, \citenamefont {Zhang},
  \citenamefont {Wang}, \citenamefont {Chen}, \citenamefont {Wu}, \citenamefont
  {Liang}, \citenamefont {Mei}, \citenamefont {Wu} \emph
  {et~al.}}]{yi2023topological}%
  \BibitemOpen
  \bibfield  {author} {\bibinfo {author} {\bibfnamefont {E.}~\bibnamefont
  {Yi}}, \bibinfo {author} {\bibfnamefont {D.~F.}\ \bibnamefont {Zheng}},
  \bibinfo {author} {\bibfnamefont {F.}~\bibnamefont {Pan}}, \bibinfo {author}
  {\bibfnamefont {H.}~\bibnamefont {Zhang}}, \bibinfo {author} {\bibfnamefont
  {B.}~\bibnamefont {Wang}}, \bibinfo {author} {\bibfnamefont {B.}~\bibnamefont
  {Chen}}, \bibinfo {author} {\bibfnamefont {D.}~\bibnamefont {Wu}}, \bibinfo
  {author} {\bibfnamefont {H.}~\bibnamefont {Liang}}, \bibinfo {author}
  {\bibfnamefont {Z.~X.}\ \bibnamefont {Mei}}, \bibinfo {author} {\bibfnamefont
  {H.}~\bibnamefont {Wu}}, \emph {et~al.},\ }\bibfield  {title} {\bibinfo
  {title} {{Topological Hall effect driven by short-range magnetic order in
  EuZn$_2$As$_2$}},\ }\href {https://doi.org/10.1103/PhysRevB.107.035142}
  {\bibfield  {journal} {\bibinfo  {journal} {Phys. Rev. B}\ }\textbf {\bibinfo
  {volume} {107}},\ \bibinfo {pages} {035142} (\bibinfo {year}
  {2023})}\BibitemShut {NoStop}%
\bibitem [{\citenamefont {Luo}\ \emph {et~al.}(2023)\citenamefont {Luo},
  \citenamefont {Xu}, \citenamefont {Du}, \citenamefont {Yang}, \citenamefont
  {Chen}, \citenamefont {Cao}, \citenamefont {Song},\ and\ \citenamefont
  {Yuan}}]{luo2023colossal}%
  \BibitemOpen
  \bibfield  {author} {\bibinfo {author} {\bibfnamefont {S.}~\bibnamefont
  {Luo}}, \bibinfo {author} {\bibfnamefont {Y.}~\bibnamefont {Xu}}, \bibinfo
  {author} {\bibfnamefont {F.}~\bibnamefont {Du}}, \bibinfo {author}
  {\bibfnamefont {L.}~\bibnamefont {Yang}}, \bibinfo {author} {\bibfnamefont
  {Y.}~\bibnamefont {Chen}}, \bibinfo {author} {\bibfnamefont {C.}~\bibnamefont
  {Cao}}, \bibinfo {author} {\bibfnamefont {Y.}~\bibnamefont {Song}},\ and\
  \bibinfo {author} {\bibfnamefont {H.}~\bibnamefont {Yuan}},\ }\bibfield
  {title} {\bibinfo {title} {{Colossal magnetoresistance and topological phase
  transition in EuZn$_2$As$_2$}},\ }\href
  {https://doi.org/10.1103/PhysRevB.108.205140} {\bibfield  {journal} {\bibinfo
   {journal} {Phys. Rev. B}\ }\textbf {\bibinfo {volume} {108}},\ \bibinfo
  {pages} {205140} (\bibinfo {year} {2023})}\BibitemShut {NoStop}%
\bibitem [{\citenamefont {Blawat}\ \emph {et~al.}(2023)\citenamefont {Blawat},
  \citenamefont {Speer}, \citenamefont {Singleton}, \citenamefont {Xie},\ and\
  \citenamefont {Jin}}]{blawat2023quantum}%
  \BibitemOpen
  \bibfield  {author} {\bibinfo {author} {\bibfnamefont {J.}~\bibnamefont
  {Blawat}}, \bibinfo {author} {\bibfnamefont {S.}~\bibnamefont {Speer}},
  \bibinfo {author} {\bibfnamefont {J.}~\bibnamefont {Singleton}}, \bibinfo
  {author} {\bibfnamefont {W.}~\bibnamefont {Xie}},\ and\ \bibinfo {author}
  {\bibfnamefont {R.}~\bibnamefont {Jin}},\ }\bibfield  {title} {\bibinfo
  {title} {{Quantum-limit phenomena and band structure in the magnetic
  topological semimetal EuZn$_2$As$_2$}},\ }\href
  {https://doi.org/10.1038/s42005-023-01378-8} {\bibfield  {journal} {\bibinfo
  {journal} {Commun. Phys.}\ }\textbf {\bibinfo {volume} {6}},\ \bibinfo
  {pages} {255} (\bibinfo {year} {2023})}\BibitemShut {NoStop}%
\bibitem [{Note1()}]{Note1}%
  \BibitemOpen
  \bibinfo {note} {See Supplemental Material at \protect \url
  {http://link.aps.org/supplemental/xxx} for the details of sample synthesis,
  experimental methods, and DFT calculations, which includes Refs.~\cite
  {homes1993technique,dressel2002electrodynamics,VASP_PW,VASP_ultrasoft,VASP_PAW,VASP_PBE,VASP_Hubbard_U,optical-kubo,Wannier90,indicator_mag}.}\BibitemShut
  {Stop}%
\bibitem [{\citenamefont {Homes}\ \emph {et~al.}(1993)\citenamefont {Homes},
  \citenamefont {Reedyk}, \citenamefont {Cradles},\ and\ \citenamefont
  {Timusk}}]{homes1993technique}%
  \BibitemOpen
  \bibfield  {author} {\bibinfo {author} {\bibfnamefont {C.~C.}\ \bibnamefont
  {Homes}}, \bibinfo {author} {\bibfnamefont {M.}~\bibnamefont {Reedyk}},
  \bibinfo {author} {\bibfnamefont {D.}~\bibnamefont {Cradles}},\ and\ \bibinfo
  {author} {\bibfnamefont {T.}~\bibnamefont {Timusk}},\ }\bibfield  {title}
  {\bibinfo {title} {Technique for measuring the reflectance of irregular,
  submillimeter-sized samples},\ }\href {https://doi.org/10.1364/AO.32.002976}
  {\bibfield  {journal} {\bibinfo  {journal} {Appl. Opt.}\ }\textbf {\bibinfo
  {volume} {32}},\ \bibinfo {pages} {2976} (\bibinfo {year}
  {1993})}\BibitemShut {NoStop}%
\bibitem [{\citenamefont {Dressel}\ and\ \citenamefont
  {Gr{\"u}ner}(2002)}]{dressel2002electrodynamics}%
  \BibitemOpen
  \bibfield  {author} {\bibinfo {author} {\bibfnamefont {M.}~\bibnamefont
  {Dressel}}\ and\ \bibinfo {author} {\bibfnamefont {G.}~\bibnamefont
  {Gr{\"u}ner}},\ }\href {https://doi.org/10.1017/CBO9780511606168} {\bibinfo
  {title} {Electrodynamics of solids: optical properties of electrons in
  matter}} (\bibinfo {year} {2002})\BibitemShut {NoStop}%
\bibitem [{\citenamefont {Kresse}\ and\ \citenamefont
  {Furthm\"uller}(1996)}]{VASP_PW}%
  \BibitemOpen
  \bibfield  {author} {\bibinfo {author} {\bibfnamefont {G.}~\bibnamefont
  {Kresse}}\ and\ \bibinfo {author} {\bibfnamefont {J.}~\bibnamefont
  {Furthm\"uller}},\ }\bibfield  {title} {\bibinfo {title} {Efficient iterative
  schemes for ab initio total-energy calculations using a plane-wave basis
  set},\ }\href {https://doi.org/10.1103/PhysRevB.54.11169} {\bibfield
  {journal} {\bibinfo  {journal} {Phys. Rev. B}\ }\textbf {\bibinfo {volume}
  {54}},\ \bibinfo {pages} {11169} (\bibinfo {year} {1996})}\BibitemShut
  {NoStop}%
\bibitem [{\citenamefont {Kresse}\ and\ \citenamefont
  {Joubert}(1999)}]{VASP_ultrasoft}%
  \BibitemOpen
  \bibfield  {author} {\bibinfo {author} {\bibfnamefont {G.}~\bibnamefont
  {Kresse}}\ and\ \bibinfo {author} {\bibfnamefont {D.}~\bibnamefont
  {Joubert}},\ }\bibfield  {title} {\bibinfo {title} {From ultrasoft
  pseudopotentials to the projector augmented-wave method},\ }\href
  {https://doi.org/10.1103/PhysRevB.59.1758} {\bibfield  {journal} {\bibinfo
  {journal} {Phys. Rev. B}\ }\textbf {\bibinfo {volume} {59}},\ \bibinfo
  {pages} {1758} (\bibinfo {year} {1999})}\BibitemShut {NoStop}%
\bibitem [{\citenamefont {Bl\"ochl}(1994)}]{VASP_PAW}%
  \BibitemOpen
  \bibfield  {author} {\bibinfo {author} {\bibfnamefont {P.~E.}\ \bibnamefont
  {Bl\"ochl}},\ }\bibfield  {title} {\bibinfo {title} {Projector augmented-wave
  method},\ }\href {https://doi.org/10.1103/PhysRevB.50.17953} {\bibfield
  {journal} {\bibinfo  {journal} {Phys. Rev. B}\ }\textbf {\bibinfo {volume}
  {50}},\ \bibinfo {pages} {17953} (\bibinfo {year} {1994})}\BibitemShut
  {NoStop}%
\bibitem [{\citenamefont {Perdew}\ \emph {et~al.}(1996)\citenamefont {Perdew},
  \citenamefont {Burke},\ and\ \citenamefont {Ernzerhof}}]{VASP_PBE}%
  \BibitemOpen
  \bibfield  {author} {\bibinfo {author} {\bibfnamefont {J.~P.}\ \bibnamefont
  {Perdew}}, \bibinfo {author} {\bibfnamefont {K.}~\bibnamefont {Burke}},\ and\
  \bibinfo {author} {\bibfnamefont {M.}~\bibnamefont {Ernzerhof}},\ }\bibfield
  {title} {\bibinfo {title} {Generalized gradient approximation made simple},\
  }\href {https://doi.org/10.1103/PhysRevLett.77.3865} {\bibfield  {journal}
  {\bibinfo  {journal} {Phys. Rev. Lett.}\ }\textbf {\bibinfo {volume} {77}},\
  \bibinfo {pages} {3865} (\bibinfo {year} {1996})}\BibitemShut {NoStop}%
\bibitem [{\citenamefont {Dudarev}\ \emph {et~al.}(1998)\citenamefont
  {Dudarev}, \citenamefont {Botton}, \citenamefont {Savrasov}, \citenamefont
  {Humphreys},\ and\ \citenamefont {Sutton}}]{VASP_Hubbard_U}%
  \BibitemOpen
  \bibfield  {author} {\bibinfo {author} {\bibfnamefont {S.~L.}\ \bibnamefont
  {Dudarev}}, \bibinfo {author} {\bibfnamefont {G.~A.}\ \bibnamefont {Botton}},
  \bibinfo {author} {\bibfnamefont {S.~Y.}\ \bibnamefont {Savrasov}}, \bibinfo
  {author} {\bibfnamefont {C.~J.}\ \bibnamefont {Humphreys}},\ and\ \bibinfo
  {author} {\bibfnamefont {A.~P.}\ \bibnamefont {Sutton}},\ }\bibfield  {title}
  {\bibinfo {title} {Electron-energy-loss spectra and the structural stability
  of nickel oxide: An lsda+u study},\ }\href
  {https://doi.org/10.1103/PhysRevB.57.1505} {\bibfield  {journal} {\bibinfo
  {journal} {Phys. Rev. B}\ }\textbf {\bibinfo {volume} {57}},\ \bibinfo
  {pages} {1505} (\bibinfo {year} {1998})}\BibitemShut {NoStop}%
\bibitem [{\citenamefont {Wang}\ and\ \citenamefont
  {Callaway}(1974)}]{optical-kubo}%
  \BibitemOpen
  \bibfield  {author} {\bibinfo {author} {\bibfnamefont {C.~S.}\ \bibnamefont
  {Wang}}\ and\ \bibinfo {author} {\bibfnamefont {J.}~\bibnamefont
  {Callaway}},\ }\bibfield  {title} {\bibinfo {title} {Band structure of
  nickel: Spin-orbit coupling, the fermi surface, and the optical
  conductivity},\ }\href {https://doi.org/10.1103/PhysRevB.9.4897} {\bibfield
  {journal} {\bibinfo  {journal} {Phys. Rev. B}\ }\textbf {\bibinfo {volume}
  {9}},\ \bibinfo {pages} {4897} (\bibinfo {year} {1974})}\BibitemShut
  {NoStop}%
\bibitem [{\citenamefont {Mostofi}\ \emph {et~al.}(2014)\citenamefont
  {Mostofi}, \citenamefont {Yates}, \citenamefont {Pizzi}, \citenamefont {Lee},
  \citenamefont {Souza}, \citenamefont {Vanderbilt},\ and\ \citenamefont
  {Marzari}}]{Wannier90}%
  \BibitemOpen
  \bibfield  {author} {\bibinfo {author} {\bibfnamefont {A.~A.}\ \bibnamefont
  {Mostofi}}, \bibinfo {author} {\bibfnamefont {J.~R.}\ \bibnamefont {Yates}},
  \bibinfo {author} {\bibfnamefont {G.}~\bibnamefont {Pizzi}}, \bibinfo
  {author} {\bibfnamefont {Y.-S.}\ \bibnamefont {Lee}}, \bibinfo {author}
  {\bibfnamefont {I.}~\bibnamefont {Souza}}, \bibinfo {author} {\bibfnamefont
  {D.}~\bibnamefont {Vanderbilt}},\ and\ \bibinfo {author} {\bibfnamefont
  {N.}~\bibnamefont {Marzari}},\ }\bibfield  {title} {\bibinfo {title} {An
  updated version of wannier90: A tool for obtaining maximally-localised
  wannier functions},\ }\href
  {https://doi.org/https://doi.org/10.1016/j.cpc.2014.05.003} {\bibfield
  {journal} {\bibinfo  {journal} {Comput. Phys. Commun.}\ }\textbf {\bibinfo
  {volume} {185}},\ \bibinfo {pages} {2309} (\bibinfo {year}
  {2014})}\BibitemShut {NoStop}%
\bibitem [{\citenamefont {Peng}\ \emph {et~al.}(2022)\citenamefont {Peng},
  \citenamefont {Jiang}, \citenamefont {Fang}, \citenamefont {Weng},\ and\
  \citenamefont {Fang}}]{indicator_mag}%
  \BibitemOpen
  \bibfield  {author} {\bibinfo {author} {\bibfnamefont {B.}~\bibnamefont
  {Peng}}, \bibinfo {author} {\bibfnamefont {Y.}~\bibnamefont {Jiang}},
  \bibinfo {author} {\bibfnamefont {Z.}~\bibnamefont {Fang}}, \bibinfo {author}
  {\bibfnamefont {H.}~\bibnamefont {Weng}},\ and\ \bibinfo {author}
  {\bibfnamefont {C.}~\bibnamefont {Fang}},\ }\bibfield  {title} {\bibinfo
  {title} {Topological classification and diagnosis in magnetically ordered
  electronic materials},\ }\href {https://doi.org/10.1103/PhysRevB.105.235138}
  {\bibfield  {journal} {\bibinfo  {journal} {Phys. Rev. B}\ }\textbf {\bibinfo
  {volume} {105}},\ \bibinfo {pages} {235138} (\bibinfo {year}
  {2022})}\BibitemShut {NoStop}%
\bibitem [{\citenamefont {Varshni}(1967)}]{varshni1967temperature}%
  \BibitemOpen
  \bibfield  {author} {\bibinfo {author} {\bibfnamefont {Y.~P.}\ \bibnamefont
  {Varshni}},\ }\bibfield  {title} {\bibinfo {title} {Temperature dependence of
  the energy gap in semiconductors},\ }\href
  {https://doi.org/10.1016/0031-8914(67)90062-6} {\bibfield  {journal}
  {\bibinfo  {journal} {Physica}\ }\textbf {\bibinfo {volume} {34}},\ \bibinfo
  {pages} {149} (\bibinfo {year} {1967})}\BibitemShut {NoStop}%
\bibitem [{\citenamefont {Homes}\ \emph {et~al.}(2023)\citenamefont {Homes},
  \citenamefont {Wang}, \citenamefont {Fruhling},\ and\ \citenamefont
  {Tafti}}]{Homes2023}%
  \BibitemOpen
  \bibfield  {author} {\bibinfo {author} {\bibfnamefont {C.~C.}\ \bibnamefont
  {Homes}}, \bibinfo {author} {\bibfnamefont {Z.-C.}\ \bibnamefont {Wang}},
  \bibinfo {author} {\bibfnamefont {K.}~\bibnamefont {Fruhling}},\ and\
  \bibinfo {author} {\bibfnamefont {F.}~\bibnamefont {Tafti}},\ }\bibfield
  {title} {\bibinfo {title} {{Optical properties and carrier localization in
  the layered phosphide ${\mathrm{EuCd}}_{2}{\mathrm{P}}_{2}$}},\ }\href
  {https://doi.org/10.1103/PhysRevB.107.045106} {\bibfield  {journal} {\bibinfo
   {journal} {Phys. Rev. B}\ }\textbf {\bibinfo {volume} {107}},\ \bibinfo
  {pages} {045106} (\bibinfo {year} {2023})}\BibitemShut {NoStop}%
\bibitem [{\citenamefont {Otrokov}\ \emph {et~al.}(2019)\citenamefont
  {Otrokov}, \citenamefont {Klimovskikh}, \citenamefont {Bentmann},
  \citenamefont {Estyunin}, \citenamefont {Zeugner}, \citenamefont {Aliev},
  \citenamefont {Ga{\ss}}, \citenamefont {Wolter}, \citenamefont {Koroleva},
  \citenamefont {Shikin} \emph {et~al.}}]{otrokov2019prediction}%
  \BibitemOpen
  \bibfield  {author} {\bibinfo {author} {\bibfnamefont {M.~M.}\ \bibnamefont
  {Otrokov}}, \bibinfo {author} {\bibfnamefont {I.~I.}\ \bibnamefont
  {Klimovskikh}}, \bibinfo {author} {\bibfnamefont {H.}~\bibnamefont
  {Bentmann}}, \bibinfo {author} {\bibfnamefont {D.}~\bibnamefont {Estyunin}},
  \bibinfo {author} {\bibfnamefont {A.}~\bibnamefont {Zeugner}}, \bibinfo
  {author} {\bibfnamefont {Z.~S.}\ \bibnamefont {Aliev}}, \bibinfo {author}
  {\bibfnamefont {S.}~\bibnamefont {Ga{\ss}}}, \bibinfo {author} {\bibfnamefont
  {A.}~\bibnamefont {Wolter}}, \bibinfo {author} {\bibfnamefont
  {A.}~\bibnamefont {Koroleva}}, \bibinfo {author} {\bibfnamefont {A.~M.}\
  \bibnamefont {Shikin}}, \emph {et~al.},\ }\bibfield  {title} {\bibinfo
  {title} {Prediction and observation of an antiferromagnetic topological
  insulator},\ }\href {https://doi.org/10.1038/s41586-019-1840-9} {\bibfield
  {journal} {\bibinfo  {journal} {Nature}\ }\textbf {\bibinfo {volume} {576}},\
  \bibinfo {pages} {416} (\bibinfo {year} {2019})}\BibitemShut {NoStop}%
\bibitem [{\citenamefont {Liu}\ \emph {et~al.}(2020)\citenamefont {Liu},
  \citenamefont {Wang}, \citenamefont {Li}, \citenamefont {Wu}, \citenamefont
  {Li}, \citenamefont {Li}, \citenamefont {He}, \citenamefont {Xu},
  \citenamefont {Zhang},\ and\ \citenamefont {Wang}}]{liu2020robust}%
  \BibitemOpen
  \bibfield  {author} {\bibinfo {author} {\bibfnamefont {C.}~\bibnamefont
  {Liu}}, \bibinfo {author} {\bibfnamefont {Y.}~\bibnamefont {Wang}}, \bibinfo
  {author} {\bibfnamefont {H.}~\bibnamefont {Li}}, \bibinfo {author}
  {\bibfnamefont {Y.}~\bibnamefont {Wu}}, \bibinfo {author} {\bibfnamefont
  {Y.}~\bibnamefont {Li}}, \bibinfo {author} {\bibfnamefont {J.}~\bibnamefont
  {Li}}, \bibinfo {author} {\bibfnamefont {K.}~\bibnamefont {He}}, \bibinfo
  {author} {\bibfnamefont {Y.}~\bibnamefont {Xu}}, \bibinfo {author}
  {\bibfnamefont {J.}~\bibnamefont {Zhang}},\ and\ \bibinfo {author}
  {\bibfnamefont {Y.}~\bibnamefont {Wang}},\ }\bibfield  {title} {\bibinfo
  {title} {Robust axion insulator and chern insulator phases in a
  two-dimensional antiferromagnetic topological insulator},\ }\href
  {https://doi.org/10.1038/s41563-019-0573-3} {\bibfield  {journal} {\bibinfo
  {journal} {Nat. Mater.}\ }\textbf {\bibinfo {volume} {19}},\ \bibinfo {pages}
  {522} (\bibinfo {year} {2020})}\BibitemShut {NoStop}%
\bibitem [{\citenamefont {Deng}\ \emph {et~al.}(2020)\citenamefont {Deng},
  \citenamefont {Yu}, \citenamefont {Shi}, \citenamefont {Guo}, \citenamefont
  {Xu}, \citenamefont {Wang}, \citenamefont {Chen},\ and\ \citenamefont
  {Zhang}}]{deng2020quantum}%
  \BibitemOpen
  \bibfield  {author} {\bibinfo {author} {\bibfnamefont {Y.}~\bibnamefont
  {Deng}}, \bibinfo {author} {\bibfnamefont {Y.}~\bibnamefont {Yu}}, \bibinfo
  {author} {\bibfnamefont {M.~Z.}\ \bibnamefont {Shi}}, \bibinfo {author}
  {\bibfnamefont {Z.}~\bibnamefont {Guo}}, \bibinfo {author} {\bibfnamefont
  {Z.}~\bibnamefont {Xu}}, \bibinfo {author} {\bibfnamefont {J.}~\bibnamefont
  {Wang}}, \bibinfo {author} {\bibfnamefont {X.~H.}\ \bibnamefont {Chen}},\
  and\ \bibinfo {author} {\bibfnamefont {Y.}~\bibnamefont {Zhang}},\ }\bibfield
   {title} {\bibinfo {title} {{Quantum anomalous Hall effect in intrinsic
  magnetic topological insulator MnBi$_2$Te$_4$}},\ }\href
  {https://doi.org/10.1126/science.aax8156} {\bibfield  {journal} {\bibinfo
  {journal} {Science}\ }\textbf {\bibinfo {volume} {367}},\ \bibinfo {pages}
  {895} (\bibinfo {year} {2020})}\BibitemShut {NoStop}%
\bibitem [{\citenamefont {Li}\ \emph {et~al.}(2019{\natexlab{b}})\citenamefont
  {Li}, \citenamefont {Li}, \citenamefont {Du}, \citenamefont {Wang},
  \citenamefont {Gu}, \citenamefont {Zhang}, \citenamefont {He}, \citenamefont
  {Duan},\ and\ \citenamefont {Xu}}]{li2019intrinsic}%
  \BibitemOpen
  \bibfield  {author} {\bibinfo {author} {\bibfnamefont {J.}~\bibnamefont
  {Li}}, \bibinfo {author} {\bibfnamefont {Y.}~\bibnamefont {Li}}, \bibinfo
  {author} {\bibfnamefont {S.}~\bibnamefont {Du}}, \bibinfo {author}
  {\bibfnamefont {Z.}~\bibnamefont {Wang}}, \bibinfo {author} {\bibfnamefont
  {B.-L.}\ \bibnamefont {Gu}}, \bibinfo {author} {\bibfnamefont {S.-C.}\
  \bibnamefont {Zhang}}, \bibinfo {author} {\bibfnamefont {K.}~\bibnamefont
  {He}}, \bibinfo {author} {\bibfnamefont {W.}~\bibnamefont {Duan}},\ and\
  \bibinfo {author} {\bibfnamefont {Y.}~\bibnamefont {Xu}},\ }\bibfield
  {title} {\bibinfo {title} {{Intrinsic magnetic topological insulators in van
  der Waals layered MnBi$_2$Te$_4$-family materials}},\ }\href
  {https://doi.org/10.1126/sciadv.aaw5685} {\bibfield  {journal} {\bibinfo
  {journal} {Sci. Adv.}\ }\textbf {\bibinfo {volume} {5}},\ \bibinfo {pages}
  {eaaw5685} (\bibinfo {year} {2019}{\natexlab{b}})}\BibitemShut {NoStop}%
\bibitem [{\citenamefont {Wang}\ \emph {et~al.}(2018)\citenamefont {Wang},
  \citenamefont {Xu}, \citenamefont {Lou}, \citenamefont {Liu}, \citenamefont
  {Li}, \citenamefont {Huang}, \citenamefont {Shen}, \citenamefont {Weng},
  \citenamefont {Wang},\ and\ \citenamefont {Lei}}]{wang2018large}%
  \BibitemOpen
  \bibfield  {author} {\bibinfo {author} {\bibfnamefont {Q.}~\bibnamefont
  {Wang}}, \bibinfo {author} {\bibfnamefont {Y.}~\bibnamefont {Xu}}, \bibinfo
  {author} {\bibfnamefont {R.}~\bibnamefont {Lou}}, \bibinfo {author}
  {\bibfnamefont {Z.}~\bibnamefont {Liu}}, \bibinfo {author} {\bibfnamefont
  {M.}~\bibnamefont {Li}}, \bibinfo {author} {\bibfnamefont {Y.}~\bibnamefont
  {Huang}}, \bibinfo {author} {\bibfnamefont {D.}~\bibnamefont {Shen}},
  \bibinfo {author} {\bibfnamefont {H.}~\bibnamefont {Weng}}, \bibinfo {author}
  {\bibfnamefont {S.}~\bibnamefont {Wang}},\ and\ \bibinfo {author}
  {\bibfnamefont {H.}~\bibnamefont {Lei}},\ }\bibfield  {title} {\bibinfo
  {title} {{Large intrinsic anomalous Hall effect in half-metallic ferromagnet
  Co$_3$Sn$_2$S$_2$ with magnetic Weyl fermions}},\ }\href
  {https://doi.org/10.1038/s41467-018-06088-2} {\bibfield  {journal} {\bibinfo
  {journal} {Nat. Commun.}\ }\textbf {\bibinfo {volume} {9}},\ \bibinfo {pages}
  {3681} (\bibinfo {year} {2018})}\BibitemShut {NoStop}%
\bibitem [{\citenamefont {Yang}\ \emph {et~al.}(2020)\citenamefont {Yang},
  \citenamefont {Zhang}, \citenamefont {Zhou}, \citenamefont {Dai},
  \citenamefont {Liao}, \citenamefont {Weng},\ and\ \citenamefont
  {Qiu}}]{yang2020magnetization}%
  \BibitemOpen
  \bibfield  {author} {\bibinfo {author} {\bibfnamefont {R.}~\bibnamefont
  {Yang}}, \bibinfo {author} {\bibfnamefont {T.}~\bibnamefont {Zhang}},
  \bibinfo {author} {\bibfnamefont {L.}~\bibnamefont {Zhou}}, \bibinfo {author}
  {\bibfnamefont {Y.}~\bibnamefont {Dai}}, \bibinfo {author} {\bibfnamefont
  {Z.}~\bibnamefont {Liao}}, \bibinfo {author} {\bibfnamefont {H.}~\bibnamefont
  {Weng}},\ and\ \bibinfo {author} {\bibfnamefont {X.}~\bibnamefont {Qiu}},\
  }\bibfield  {title} {\bibinfo {title} {{Magnetization-induced band shift in
  ferromagnetic Weyl semimetal Co$_3$Sn$_2$S$_2$}},\ }\href
  {https://doi.org/10.1103/PhysRevLett.124.077403} {\bibfield  {journal}
  {\bibinfo  {journal} {Phys. Rev. Lett.}\ }\textbf {\bibinfo {volume} {124}},\
  \bibinfo {pages} {077403} (\bibinfo {year} {2020})}\BibitemShut {NoStop}%
\bibitem [{\citenamefont {Santos-Cottin}\ \emph {et~al.}(2023)\citenamefont
  {Santos-Cottin}, \citenamefont {Mohelsk{\`y}}, \citenamefont {Wyzula},
  \citenamefont {Le~Mardel{\'e}}, \citenamefont {Kapon}, \citenamefont
  {Nasrallah}, \citenamefont {Bari{\v{s}}i{\'c}}, \citenamefont
  {{\v{Z}}ivkovi{\'c}}, \citenamefont {Soh}, \citenamefont {Guo} \emph
  {et~al.}}]{santos2023eucd}%
  \BibitemOpen
  \bibfield  {author} {\bibinfo {author} {\bibfnamefont {D.}~\bibnamefont
  {Santos-Cottin}}, \bibinfo {author} {\bibfnamefont {I.}~\bibnamefont
  {Mohelsk{\`y}}}, \bibinfo {author} {\bibfnamefont {J.}~\bibnamefont
  {Wyzula}}, \bibinfo {author} {\bibfnamefont {F.}~\bibnamefont
  {Le~Mardel{\'e}}}, \bibinfo {author} {\bibfnamefont {I.}~\bibnamefont
  {Kapon}}, \bibinfo {author} {\bibfnamefont {S.}~\bibnamefont {Nasrallah}},
  \bibinfo {author} {\bibfnamefont {N.}~\bibnamefont {Bari{\v{s}}i{\'c}}},
  \bibinfo {author} {\bibfnamefont {I.}~\bibnamefont {{\v{Z}}ivkovi{\'c}}},
  \bibinfo {author} {\bibfnamefont {J.}~\bibnamefont {Soh}}, \bibinfo {author}
  {\bibfnamefont {F.}~\bibnamefont {Guo}}, \emph {et~al.},\ }\bibfield  {title}
  {\bibinfo {title} {{EuCd$_2$As$_2$: A Magnetic Semiconductor}},\ }\href
  {https://doi.org/10.1103/PhysRevLett.131.186704} {\bibfield  {journal}
  {\bibinfo  {journal} {Phys. Rev. Lett.}\ }\textbf {\bibinfo {volume} {131}},\
  \bibinfo {pages} {186704} (\bibinfo {year} {2023})}\BibitemShut {NoStop}%
\end{thebibliography}

\begin{thebibliography}{13}%
\makeatletter
\providecommand \@ifxundefined [1]{%
 \@ifx{#1\undefined}
}%
\providecommand \@ifnum [1]{%
 \ifnum #1\expandafter \@firstoftwo
 \else \expandafter \@secondoftwo
 \fi
}%
\providecommand \@ifx [1]{%
 \ifx #1\expandafter \@firstoftwo
 \else \expandafter \@secondoftwo
 \fi
}%
\providecommand \natexlab [1]{#1}%
\providecommand \enquote  [1]{``#1''}%
\providecommand \bibnamefont  [1]{#1}%
\providecommand \bibfnamefont [1]{#1}%
\providecommand \citenamefont [1]{#1}%
\providecommand \href@noop [0]{\@secondoftwo}%
\providecommand \href [0]{\begingroup \@sanitize@url \@href}%
\providecommand \@href[1]{\@@startlink{#1}\@@href}%
\providecommand \@@href[1]{\endgroup#1\@@endlink}%
\providecommand \@sanitize@url [0]{\catcode `\\12\catcode `\$12\catcode
  `\&12\catcode `\#12\catcode `\^12\catcode `\_12\catcode `\%12\relax}%
\providecommand \@@startlink[1]{}%
\providecommand \@@endlink[0]{}%
\providecommand \url  [0]{\begingroup\@sanitize@url \@url }%
\providecommand \@url [1]{\endgroup\@href {#1}{\urlprefix }}%
\providecommand \urlprefix  [0]{URL }%
\providecommand \Eprint [0]{\href }%
\providecommand \doibase [0]{https://doi.org/}%
\providecommand \selectlanguage [0]{\@gobble}%
\providecommand \bibinfo  [0]{\@secondoftwo}%
\providecommand \bibfield  [0]{\@secondoftwo}%
\providecommand \translation [1]{[#1]}%
\providecommand \BibitemOpen [0]{}%
\providecommand \bibitemStop [0]{}%
\providecommand \bibitemNoStop [0]{.\EOS\space}%
\providecommand \EOS [0]{\spacefactor3000\relax}%
\providecommand \BibitemShut  [1]{\csname bibitem#1\endcsname}%
\let\auto@bib@innerbib\@empty
%</preamble>
\bibitem [{\citenamefont {Wang}\ \emph {et~al.}(2022)\citenamefont {Wang},
  \citenamefont {Been}, \citenamefont {Gaudet}, \citenamefont {Alqasseri},
  \citenamefont {Fruhling}, \citenamefont {Yao}, \citenamefont {Stuhr},
  \citenamefont {Zhu}, \citenamefont {Ren}, \citenamefont {Cui} \emph
  {et~al.}}]{wang2022anisotropy}%
  \BibitemOpen
  \bibfield  {author} {\bibinfo {author} {\bibfnamefont {Z.-C.}\ \bibnamefont
  {Wang}}, \bibinfo {author} {\bibfnamefont {E.}~\bibnamefont {Been}}, \bibinfo
  {author} {\bibfnamefont {J.}~\bibnamefont {Gaudet}}, \bibinfo {author}
  {\bibfnamefont {G.~M.~A.}\ \bibnamefont {Alqasseri}}, \bibinfo {author}
  {\bibfnamefont {K.}~\bibnamefont {Fruhling}}, \bibinfo {author}
  {\bibfnamefont {X.}~\bibnamefont {Yao}}, \bibinfo {author} {\bibfnamefont
  {U.}~\bibnamefont {Stuhr}}, \bibinfo {author} {\bibfnamefont
  {Q.}~\bibnamefont {Zhu}}, \bibinfo {author} {\bibfnamefont {Z.}~\bibnamefont
  {Ren}}, \bibinfo {author} {\bibfnamefont {Y.}~\bibnamefont {Cui}}, \emph
  {et~al.},\ }\bibfield  {title} {\bibinfo {title} {{Anisotropy of the magnetic
  and transport properties of EuZn$_2$As$_2$}},\ }\href
  {https://doi.org/10.1103/PhysRevB.105.165122} {\bibfield  {journal} {\bibinfo
   {journal} {Phys. Rev. B}\ }\textbf {\bibinfo {volume} {105}},\ \bibinfo
  {pages} {165122} (\bibinfo {year} {2022})}\BibitemShut {NoStop}%
\bibitem [{\citenamefont {Blawat}\ \emph {et~al.}(2022)\citenamefont {Blawat},
  \citenamefont {Marshall}, \citenamefont {Singleton}, \citenamefont {Feng},
  \citenamefont {Cao}, \citenamefont {Xie},\ and\ \citenamefont
  {Jin}}]{blawat2022unusual}%
  \BibitemOpen
  \bibfield  {author} {\bibinfo {author} {\bibfnamefont {J.}~\bibnamefont
  {Blawat}}, \bibinfo {author} {\bibfnamefont {M.}~\bibnamefont {Marshall}},
  \bibinfo {author} {\bibfnamefont {J.}~\bibnamefont {Singleton}}, \bibinfo
  {author} {\bibfnamefont {E.}~\bibnamefont {Feng}}, \bibinfo {author}
  {\bibfnamefont {H.}~\bibnamefont {Cao}}, \bibinfo {author} {\bibfnamefont
  {W.}~\bibnamefont {Xie}},\ and\ \bibinfo {author} {\bibfnamefont
  {R.}~\bibnamefont {Jin}},\ }\bibfield  {title} {\bibinfo {title} {{Unusual
  electrical and magnetic properties in layered EuZn$_2$As$_2$}},\ }\href
  {https://doi.org/10.1002/qute.202200012} {\bibfield  {journal} {\bibinfo
  {journal} {Adv. Quantum Technol.}\ }\textbf {\bibinfo {volume} {5}},\
  \bibinfo {pages} {2200012} (\bibinfo {year} {2022})}\BibitemShut {NoStop}%
\bibitem [{\citenamefont {Bukowski}\ \emph {et~al.}(2022)\citenamefont
  {Bukowski}, \citenamefont {Rybicki}, \citenamefont {Babij}, \citenamefont
  {Przewo{\'z}nik}, \citenamefont {Gondek}, \citenamefont {{\.Z}ukrowski},\
  and\ \citenamefont {Kapusta}}]{bukowski2022canted}%
  \BibitemOpen
  \bibfield  {author} {\bibinfo {author} {\bibfnamefont {Z.}~\bibnamefont
  {Bukowski}}, \bibinfo {author} {\bibfnamefont {D.}~\bibnamefont {Rybicki}},
  \bibinfo {author} {\bibfnamefont {M.}~\bibnamefont {Babij}}, \bibinfo
  {author} {\bibfnamefont {J.}~\bibnamefont {Przewo{\'z}nik}}, \bibinfo
  {author} {\bibfnamefont {{\L}.}~\bibnamefont {Gondek}}, \bibinfo {author}
  {\bibfnamefont {J.}~\bibnamefont {{\.Z}ukrowski}},\ and\ \bibinfo {author}
  {\bibfnamefont {C.}~\bibnamefont {Kapusta}},\ }\bibfield  {title} {\bibinfo
  {title} {{Canted antiferromagnetic order in EuZn$_2$As$_2$ single
  crystals}},\ }\href {https://doi.org/10.1038/s41598-022-19026-6} {\bibfield
  {journal} {\bibinfo  {journal} {Sci. Rep.}\ }\textbf {\bibinfo {volume}
  {12}},\ \bibinfo {pages} {14718} (\bibinfo {year} {2022})}\BibitemShut
  {NoStop}%
\bibitem [{\citenamefont {Homes}\ \emph {et~al.}(1993)\citenamefont {Homes},
  \citenamefont {Reedyk}, \citenamefont {Cradles},\ and\ \citenamefont
  {Timusk}}]{homes1993technique}%
  \BibitemOpen
  \bibfield  {author} {\bibinfo {author} {\bibfnamefont {C.~C.}\ \bibnamefont
  {Homes}}, \bibinfo {author} {\bibfnamefont {M.}~\bibnamefont {Reedyk}},
  \bibinfo {author} {\bibfnamefont {D.}~\bibnamefont {Cradles}},\ and\ \bibinfo
  {author} {\bibfnamefont {T.}~\bibnamefont {Timusk}},\ }\bibfield  {title}
  {\bibinfo {title} {Technique for measuring the reflectance of irregular,
  submillimeter-sized samples},\ }\href {https://doi.org/10.1364/AO.32.002976}
  {\bibfield  {journal} {\bibinfo  {journal} {Appl. Opt.}\ }\textbf {\bibinfo
  {volume} {32}},\ \bibinfo {pages} {2976} (\bibinfo {year}
  {1993})}\BibitemShut {NoStop}%
\bibitem [{\citenamefont {Dressel}\ and\ \citenamefont
  {Gr{\"u}ner}(2002)}]{dressel2002electrodynamics}%
  \BibitemOpen
  \bibfield  {author} {\bibinfo {author} {\bibfnamefont {M.}~\bibnamefont
  {Dressel}}\ and\ \bibinfo {author} {\bibfnamefont {G.}~\bibnamefont
  {Gr{\"u}ner}},\ }\href {https://doi.org/10.1017/CBO9780511606168} {\bibinfo
  {title} {Electrodynamics of solids: optical properties of electrons in
  matter}} (\bibinfo {year} {2002})\BibitemShut {NoStop}%
\bibitem [{\citenamefont {Kresse}\ and\ \citenamefont
  {Furthm\"uller}(1996)}]{VASP_PW}%
  \BibitemOpen
  \bibfield  {author} {\bibinfo {author} {\bibfnamefont {G.}~\bibnamefont
  {Kresse}}\ and\ \bibinfo {author} {\bibfnamefont {J.}~\bibnamefont
  {Furthm\"uller}},\ }\bibfield  {title} {\bibinfo {title} {Efficient iterative
  schemes for ab initio total-energy calculations using a plane-wave basis
  set},\ }\href {https://doi.org/10.1103/PhysRevB.54.11169} {\bibfield
  {journal} {\bibinfo  {journal} {Phys. Rev. B}\ }\textbf {\bibinfo {volume}
  {54}},\ \bibinfo {pages} {11169} (\bibinfo {year} {1996})}\BibitemShut
  {NoStop}%
\bibitem [{\citenamefont {Kresse}\ and\ \citenamefont
  {Joubert}(1999)}]{VASP_ultrasoft}%
  \BibitemOpen
  \bibfield  {author} {\bibinfo {author} {\bibfnamefont {G.}~\bibnamefont
  {Kresse}}\ and\ \bibinfo {author} {\bibfnamefont {D.}~\bibnamefont
  {Joubert}},\ }\bibfield  {title} {\bibinfo {title} {From ultrasoft
  pseudopotentials to the projector augmented-wave method},\ }\href
  {https://doi.org/10.1103/PhysRevB.59.1758} {\bibfield  {journal} {\bibinfo
  {journal} {Phys. Rev. B}\ }\textbf {\bibinfo {volume} {59}},\ \bibinfo
  {pages} {1758} (\bibinfo {year} {1999})}\BibitemShut {NoStop}%
\bibitem [{\citenamefont {Bl\"ochl}(1994)}]{VASP_PAW}%
  \BibitemOpen
  \bibfield  {author} {\bibinfo {author} {\bibfnamefont {P.~E.}\ \bibnamefont
  {Bl\"ochl}},\ }\bibfield  {title} {\bibinfo {title} {Projector augmented-wave
  method},\ }\href {https://doi.org/10.1103/PhysRevB.50.17953} {\bibfield
  {journal} {\bibinfo  {journal} {Phys. Rev. B}\ }\textbf {\bibinfo {volume}
  {50}},\ \bibinfo {pages} {17953} (\bibinfo {year} {1994})}\BibitemShut
  {NoStop}%
\bibitem [{\citenamefont {Perdew}\ \emph {et~al.}(1996)\citenamefont {Perdew},
  \citenamefont {Burke},\ and\ \citenamefont {Ernzerhof}}]{VASP_PBE}%
  \BibitemOpen
  \bibfield  {author} {\bibinfo {author} {\bibfnamefont {J.~P.}\ \bibnamefont
  {Perdew}}, \bibinfo {author} {\bibfnamefont {K.}~\bibnamefont {Burke}},\ and\
  \bibinfo {author} {\bibfnamefont {M.}~\bibnamefont {Ernzerhof}},\ }\bibfield
  {title} {\bibinfo {title} {Generalized gradient approximation made simple},\
  }\href {https://doi.org/10.1103/PhysRevLett.77.3865} {\bibfield  {journal}
  {\bibinfo  {journal} {Phys. Rev. Lett.}\ }\textbf {\bibinfo {volume} {77}},\
  \bibinfo {pages} {3865} (\bibinfo {year} {1996})}\BibitemShut {NoStop}%
\bibitem [{\citenamefont {Dudarev}\ \emph {et~al.}(1998)\citenamefont
  {Dudarev}, \citenamefont {Botton}, \citenamefont {Savrasov}, \citenamefont
  {Humphreys},\ and\ \citenamefont {Sutton}}]{VASP_Hubbard_U}%
  \BibitemOpen
  \bibfield  {author} {\bibinfo {author} {\bibfnamefont {S.~L.}\ \bibnamefont
  {Dudarev}}, \bibinfo {author} {\bibfnamefont {G.~A.}\ \bibnamefont {Botton}},
  \bibinfo {author} {\bibfnamefont {S.~Y.}\ \bibnamefont {Savrasov}}, \bibinfo
  {author} {\bibfnamefont {C.~J.}\ \bibnamefont {Humphreys}},\ and\ \bibinfo
  {author} {\bibfnamefont {A.~P.}\ \bibnamefont {Sutton}},\ }\bibfield  {title}
  {\bibinfo {title} {Electron-energy-loss spectra and the structural stability
  of nickel oxide: An lsda+u study},\ }\href
  {https://doi.org/10.1103/PhysRevB.57.1505} {\bibfield  {journal} {\bibinfo
  {journal} {Phys. Rev. B}\ }\textbf {\bibinfo {volume} {57}},\ \bibinfo
  {pages} {1505} (\bibinfo {year} {1998})}\BibitemShut {NoStop}%
\bibitem [{\citenamefont {Wang}\ and\ \citenamefont
  {Callaway}(1974)}]{optical-kubo}%
  \BibitemOpen
  \bibfield  {author} {\bibinfo {author} {\bibfnamefont {C.~S.}\ \bibnamefont
  {Wang}}\ and\ \bibinfo {author} {\bibfnamefont {J.}~\bibnamefont
  {Callaway}},\ }\bibfield  {title} {\bibinfo {title} {Band structure of
  nickel: Spin-orbit coupling, the fermi surface, and the optical
  conductivity},\ }\href {https://doi.org/10.1103/PhysRevB.9.4897} {\bibfield
  {journal} {\bibinfo  {journal} {Phys. Rev. B}\ }\textbf {\bibinfo {volume}
  {9}},\ \bibinfo {pages} {4897} (\bibinfo {year} {1974})}\BibitemShut
  {NoStop}%
\bibitem [{\citenamefont {Mostofi}\ \emph {et~al.}(2014)\citenamefont
  {Mostofi}, \citenamefont {Yates}, \citenamefont {Pizzi}, \citenamefont {Lee},
  \citenamefont {Souza}, \citenamefont {Vanderbilt},\ and\ \citenamefont
  {Marzari}}]{Wannier90}%
  \BibitemOpen
  \bibfield  {author} {\bibinfo {author} {\bibfnamefont {A.~A.}\ \bibnamefont
  {Mostofi}}, \bibinfo {author} {\bibfnamefont {J.~R.}\ \bibnamefont {Yates}},
  \bibinfo {author} {\bibfnamefont {G.}~\bibnamefont {Pizzi}}, \bibinfo
  {author} {\bibfnamefont {Y.-S.}\ \bibnamefont {Lee}}, \bibinfo {author}
  {\bibfnamefont {I.}~\bibnamefont {Souza}}, \bibinfo {author} {\bibfnamefont
  {D.}~\bibnamefont {Vanderbilt}},\ and\ \bibinfo {author} {\bibfnamefont
  {N.}~\bibnamefont {Marzari}},\ }\bibfield  {title} {\bibinfo {title} {An
  updated version of wannier90: A tool for obtaining maximally-localised
  wannier functions},\ }\href
  {https://doi.org/https://doi.org/10.1016/j.cpc.2014.05.003} {\bibfield
  {journal} {\bibinfo  {journal} {Comput. Phys. Commun.}\ }\textbf {\bibinfo
  {volume} {185}},\ \bibinfo {pages} {2309} (\bibinfo {year}
  {2014})}\BibitemShut {NoStop}%
\bibitem [{\citenamefont {Peng}\ \emph {et~al.}(2022)\citenamefont {Peng},
  \citenamefont {Jiang}, \citenamefont {Fang}, \citenamefont {Weng},\ and\
  \citenamefont {Fang}}]{indicator_mag}%
  \BibitemOpen
  \bibfield  {author} {\bibinfo {author} {\bibfnamefont {B.}~\bibnamefont
  {Peng}}, \bibinfo {author} {\bibfnamefont {Y.}~\bibnamefont {Jiang}},
  \bibinfo {author} {\bibfnamefont {Z.}~\bibnamefont {Fang}}, \bibinfo {author}
  {\bibfnamefont {H.}~\bibnamefont {Weng}},\ and\ \bibinfo {author}
  {\bibfnamefont {C.}~\bibnamefont {Fang}},\ }\bibfield  {title} {\bibinfo
  {title} {Topological classification and diagnosis in magnetically ordered
  electronic materials},\ }\href {https://doi.org/10.1103/PhysRevB.105.235138}
  {\bibfield  {journal} {\bibinfo  {journal} {Phys. Rev. B}\ }\textbf {\bibinfo
  {volume} {105}},\ \bibinfo {pages} {235138} (\bibinfo {year}
  {2022})}\BibitemShut {NoStop}%
\end{thebibliography}

\end{document}